\documentclass[11pt]{article}
\RequirePackage[OT1]{fontenc} \RequirePackage{amsthm,amsmath}
\usepackage{graphics,epsfig,epsf,psfrag}
\usepackage{url}
\usepackage{amsmath,amsfonts,amssymb,amsthm}
\usepackage{bbm, bm}
\usepackage{graphicx,lscape,rotate}
\usepackage{movie15}  
\usepackage{dashrule} 
\usepackage{epstopdf} 
\usepackage{enumerate, framed}
\usepackage{colortbl,booktabs}
\usepackage{color, appendix}
\usepackage[round]{natbib}
\RequirePackage[colorlinks,citecolor=blue,urlcolor=blue]{hyperref}
\usepackage{pstricks,pst-node,pst-tree}  
\usepackage{algorithmic} 
\usepackage[]{algorithm2e}
\usepackage{dsfont,setspace,subcaption}

\makeatother

\usepackage{graphics, hyperref}  

\newtheorem{theorem}{Theorem}

\newtheorem{proposition}{Proposition}

\newtheorem*{remark*}{Remark} 
\theoremstyle{definition}
\newtheorem{assumption}{Assumption}

\newcommand{\sinc}{\mbox{sinc}}

\newcount\colveccount
\newcommand*\colvec[1]{
        \global\colveccount#1
        \begin{pmatrix}
        \colvecnext
}
\def\colvecnext#1{
        #1
        \global\advance\colveccount-1
        \ifnum\colveccount>0
                \\
                \expandafter\colvecnext
        \else
                \end{pmatrix}
        \fi
}

\makeatletter
\newcommand{\Spvek}[2][r]{%
  \gdef\@VORNE{1}
  \left(\hskip-\arraycolsep%
    \begin{array}{#1}\vekSp@lten{#2}\end{array}%
  \hskip-\arraycolsep\right)}

\def\vekSp@lten#1{\xvekSp@lten#1;vekL@stLine;}
\def\vekL@stLine{vekL@stLine}
\def\xvekSp@lten#1;{\def\temp{#1}%
  \ifx\temp\vekL@stLine
  \else
    \ifnum\@VORNE=1\gdef\@VORNE{0}
    \else\@arraycr\fi%
    #1%
    \expandafter\xvekSp@lten
  \fi}
\makeatother

\begin{document}


\title{Kernel Density Estimation and Convolution Revisited}
\author{\textbf{Nicholas Tenkorang}, 
\textbf{Kwesi Appau Ohene-Obeng}, and \textbf{Xiaogang Su} \\
Department of Mathematical Sciences \\
University of Texas, El Paso, TX 79968 \vspace{.1in} }

\date{\today}
\maketitle

\renewcommand{\abstractname}{\large Abstract \vspace{.1in}}
\begin{abstract}
{\normalsize
Kernel Density Estimation (KDE) is a cornerstone of nonparametric statistics, yet it remains sensitive to bandwidth choice, boundary bias, and computational inefficiency. This study revisits KDE through a principled convolutional framework, providing an intuitive model-based derivation that naturally extends to constrained domains, such as positive-valued random variables. Building on this perspective, we introduce SHIDE (Simulation and Histogram Interpolation for Density Estimation), a novel and computationally efficient density estimator that generates pseudo-data by adding bounded noise to observations and applies spline interpolation to the resulting histogram. The noise is sampled from a class of bounded polynomial kernel densities, constructed through convolutions of uniform distributions, with a natural bandwidth parameter defined by the kernel's support bound. We establish the theoretical properties of SHIDE, including pointwise consistency, bias-variance decomposition, and asymptotic MISE, showing that SHIDE attains the classical $n^{-4/5}$ convergence rate while mitigating boundary bias. Two data–driven bandwidth selection methods are developed, an AMISE–optimal rule and a percentile–based alternative, which are shown to be asymptotically equivalent. Extensive simulations demonstrate that SHIDE performs comparably to or surpasses KDE across a broad range of models, with particular advantages for bounded and heavy-tailed distributions. These results highlight SHIDE as a theoretically grounded and practically robust alternative to traditional KDE. }
\end{abstract}

\noindent%

\section{Introduction}
\label{sec-intro}

Kernel Density Estimation (KDE) \citet{Rosenblatt:1956, Parzen:1962} is a powerful statistical technique for estimating and visualizing the probability density functions of random variables. It serves as a versatile tool across data science, machine learning, and artificial intelligence. In density-based clustering algorithms such as mean-shift clustering \citep{Cheng:1995}, KDE identifies cluster centers by locating local maxima of the estimated density. It also underpins anomaly detection systems by flagging data points that fall in low-probability regions \citep{Scholkopf:1999}. For generative modeling, KDE offers a non-parametric approach to approximate complex data distributions and generate synthetic samples \citep{Silverman:1986}. In Bayesian inference, it is used to approximate posterior distributions and construct proposal distributions for sampling-based methods \citep{Wand:1995}. KDE further contributes to non-parametric regression by optimizing bandwidth parameters and enabling density-aware feature engineering in predictive modeling \citep{Nadaraya:1964}. In computer vision, its applications range from object tracking via mean-shift algorithms to background subtraction in video analysis \citep{Comaniciu:2002}. In natural language processing, KDE helps smooth word distributions in topic models \citep{Blei:2003}, while in reinforcement learning \citep{Sutton:2018}, it facilitates exploration of the action space. Additional applications include uncertainty quantification through confidence interval estimation \citep{Sheather:1991} and topological data analysis via density-aware filtrations \citep{Chazal:2018}. 
 
A key mathematical concept that underpins KDE is convolution, which plays a crucial role in how density estimates are constructed. Convolution is a fundamental operation in mathematics that describes how the shape of one function transforms another through a blending process. It captures how systems respond to combined inputs by integrating overlapping effects, such as smoothing a signal or modifying an image \citep{Oppenheim:1997}. In engineering, convolution enables noise reduction in audio processing and sharpens edge detection in computer vision. It also underpins modern artificial intelligence, where convolutional layers in neural networks extract hierarchical patterns from data, powering breakthroughs in image and speech recognition \citep{Goodfellow:2016}. From simulating physical phenomena to enhancing medical imaging, convolution’s ability to model interacting influences makes it a cornerstone of applied mathematics and computational science. 

In statistics, convolution is known to be the method of computing the PDF of the sum of two random variables. In this work, we will explain the relationship between KDE and convolution via a simple statistical model, making the KDE formula more accessible to students and practitioners. The similar modeling approach can be extended to special cases such as positive random variables, which leads to the multiplicative convolution \citep{{Comte:2012}}. Furthermore, we put forward a new simulation-based method for density estimation from the same model. The main idea is to generate new pseudo-data based on current observations by adding noises and smooth the histogram with splines. To this end, we introduce a class of noise variables whose densities are polynomial kernels with bounded supports. These noise varibles are obtained via iterative convolution of uniform random variables over $[-1/2, 1/2]$. The proposed method also has a natural smoothing parameter, which is the bound of the support. A practically appealing choice can be found as a percentile of the neighboring distance of sorted observations in the current data. The new approach has greater flexibility to handle special random variables, such as positive, negative, bounded, or heavy-tailed ones, simply through transformation. Numerical studies show that the proposed method can perform better or equally well than KDE in various scenarios.

The remainder of this article is organized as follows. Section~\ref{sec-convolution} reinterprets classical KDE through a convolutional framework and extends it to constrained domains such as positive or bounded supports. Building on this foundation, Section~\ref{sec-SHIDE} introduces the proposed SHIDE method, describing its pseudo-data generation scheme and spline-based histogram interpolation. Section~\ref{sec-theory} presents a comprehensive theoretical analysis, establishing SHIDE’s pointwise convergence, bias–variance decomposition, and asymptotic convergence rate, and comparing its properties with those of KDE. Section~\ref{sec-bandwidth} develops two bandwidth selection strategies, an AMISE-optimal rule and a percentile-based alternative, grounded in the theoretical framework. Section~\ref{sec-numerical} reports simulation studies that evaluate SHIDE’s performance relative to KDE across a range of distributional settings. Finally, Section~\ref{sec-discussion} concludes with a discussion of the method’s implications, limitations, practical considerations, and potential future directions.

\section{KDE and Convolution}
\label{sec-convolution}

Kernel density estimation (KDE) is a nonparametric method for constructing a smooth estimate of a probability density function from a finite sample of data.  Given a sample $\mathcal{D}$ of independent and identically distributed (IID) observations $x_1, x_2, \dots, x_n$ drawn from an unknown, smooth density $f(x)$, the KDE is defined by
\begin{equation}
\label{KDE}
\widehat{f}(x) = \frac{1}{n} \sum_{i=1}^{n} K_h(x - x_i),
\end{equation}
where $\widehat{f}(x)$ is the estimated density at the point $x$, and $K_h$ is the kernel function scaled by the bandwidth $h > 0$.  The scaled kernel is defined by
\begin{equation}
\label{scaled-kernel}
K_h(x) = \frac{1}{h} K\Bigl(\frac{x}{h}\Bigr).
\end{equation}
The kernel function $K(\cdot)$ must be integrable ($K \in L^1(\mathbb{R})$) and normalized so that $\int_{\mathbb{R}} K(x)\,dx = 1$.  Common choices of $K(\cdot)$ include the Gaussian, Epanechnikov, and uniform kernels.  The bandwidth $h$ acts as a smoothing parameter that critically affects the shape and smoothness of the estimate; many methods have been proposed for selecting $h$.

\subsection{Additive Convolution KDE}
\label{sec-additiveKDE}

Although the KDE formula in \eqref{KDE} may seem unintuitive at first glance, it can be interpreted more intuitively via the following additive model:
\begin{equation}
\label{model-symbolic}
X' = X + \varepsilon,
\end{equation}
where $X'$ denotes the true underlying random variable, $X$ is the observed variable, and $\varepsilon$ is an independent noise term ($\varepsilon \perp X$) with zero mean ($\mathbb{E}[\varepsilon] = 0$).  The noise term $\varepsilon$ can be seen as capturing measurement error or other sources of randomness.

Under the additive model \eqref{model-symbolic}, the density of \( X' \) can be expressed as the convolution of the density of \( X \) with that of \( \varepsilon \). Since \( X \) corresponds to an empirical distribution consisting of point masses located at the observed values \( x_1, x_2, \ldots, x_n \), it can be formally described using an empirical measure. In the framework of generalized functions (distributions), the corresponding formal ``density" of \( X \) is given by
\begin{equation}
\label{ePDF}
g(x) = \frac{1}{n} \sum_{i=1}^n \delta(x - x_i),
\end{equation}
where \( \delta(\cdot) \) denotes the Dirac delta function, defined by the property \( \delta(x) = 0 \) for \( x \neq 0 \) and
\(
\int_{-\infty}^{\infty} \delta(x)\, dx = 1.
\)
Although \( g(x) \) is not a classical probability density function (PDF), this representation is well-defined in the distributional sense and is valid when used in convolution with a smooth kernel (see, e.g., \cite{Silverman:1986}).

Assuming the noise variable \( \varepsilon \) has density \( K_h \), the density of \( X' \) can be written as the convolution of \( g \) and \( K_h \):
\begin{eqnarray}
\label{convolution}
\widehat{f}(x) &=& (g \star K_h)(x) = \int_{\mathbb{R}} g(x - u)\, K_h(u)\, du \nonumber \\
&=& \frac{1}{n} \sum_{i=1}^n \int_{\mathbb{R}} \delta(x - u - x_i)\, K_h(u)\, du \nonumber \\
&=& \frac{1}{n} \sum_{i=1}^n K_h(x - x_i) \qquad \text{(by the sifting property of } \delta(\cdot)\text{)} \nonumber \\
&=& \frac{1}{nh} \sum_{i=1}^n K\left( \frac{x - x_i}{h} \right), \nonumber 
\end{eqnarray}
which recovers the KDE expression in \eqref{KDE}. Here, \( \star \) denotes the convolution operator. The simplification in the third line follows from the sifting (or sampling) property of the Dirac delta function, which states that for any smooth function \( \psi(\cdot) \),
\[
(\psi \star \delta)(c) = \int_{\mathbb{R}} \psi(x)\, \delta(c - x)\, dx = \psi(c).
\]

The kernel function \(K(x)\) serves as a smooth approximation tool via convolution.  For improved approximation properties, it is customary to require \(K(x)\) to decay rapidly to zero as \(\lvert x\rvert\to\infty\), be nonnegative (\(K(x)\ge0\)), symmetric (\(K(x)=K(-x)\)), integrate to one (\(\int_{-\infty}^\infty K(x)\,dx=1\)), and have finite variance (\(\int_{-\infty}^\infty x^2\,K(x)\,dx<\infty\)) \citep{Cheney:2000}. Clearly, \(K(\cdot)\) is itself a valid PDF, and so is its scaled version $K_h(x)$, justifying the assumption \(\varepsilon\sim K_h\).  In particular, choosing \(K\) to be the Gaussian density yields the familiar Gaussian kernel.

\subsection{Multiplicative Convolution KDE for Positive/Negative Variables}
\label{sec-multiplicativeKDE}

When the data are strictly positive, a \emph{multiplicative} model offers a more natural and effective alternative:
\begin{equation}
\label{model-multiplicative}
X' = X\,\varepsilon,
\end{equation}
where \( X > 0 \) represents the observed positive variable, \( X' > 0 \) is the true latent variable, and \( \varepsilon > 0 \) is a random scale factor with density \( K_h \). Assuming \( X \sim g \) and \( \varepsilon \sim K_h \), the density of the product \( X' = X\varepsilon \) is given by the multiplicative convolution
\[
\widehat f(x)
= \int_{0}^{\infty} g\!\left(\displaystyle\frac{x}{u}\right)\,\displaystyle\frac{1}{u}\,K_h(u)\,du.
\]
Substituting $g(u)$ in (\ref{ePDF}) into the integral yields
\[
\widehat f(x)
= \int_{0}^{\infty} \left[ \displaystyle\frac{1}{n} \sum_{i=1}^n \delta\left( \displaystyle\frac{x}{u} - x_i \right) \right] \displaystyle\frac{1}{u} K_h(u)\,du
= \displaystyle\frac{1}{n} \sum_{i=1}^n \int_{0}^{\infty} \delta\left( \displaystyle\frac{x}{u} - x_i \right) \displaystyle\frac{1}{u} K_h(u)\,du.
\]
We apply the sifting property of the Dirac delta function under the change of variables \( u \mapsto x / u \):
\[
\delta\!\left( \displaystyle\frac{x}{u} - x_i \right)
= \displaystyle\frac{x}{x_i^2}\, \delta\!\left( u - \displaystyle\frac{x}{x_i} \right).
\]
The integral becomes
\[
\widehat f(x)
= \displaystyle\frac{1}{n} \sum_{i=1}^n \int_{0}^{\infty} \displaystyle\frac{x}{x_i^2}\, \delta\left( u - \displaystyle\frac{x}{x_i} \right)\, \displaystyle\frac{1}{u} K_h(u)\,du
= \displaystyle\frac{1}{n} \sum_{i=1}^n \displaystyle\frac{1}{x_i} K_h\!\left( \displaystyle\frac{x}{x_i} \right).
\]
Recalling the scaled kernel form \( K_h(u)\) in (\ref{scaled-kernel}), the final KDE for positive data is
\begin{equation}
\label{KDE-positive}
\widehat f(x)
= \displaystyle\frac{1}{n\,h} \sum_{i=1}^n \displaystyle\frac{1}{x_i}\, K\!\left( \displaystyle\frac{x}{h\,x_i} \right),
\end{equation}
which is known as the convolution power kernel density estimator, as studied in \cite{Comte:2012}.

A similar construction applies when the data are strictly negative. In that case, we may use the transformation \( X' = X \varepsilon \), where \( X < 0 \), \( X' < 0 \), and \( \varepsilon > 0 \) again serves as a positive scale factor. The only change in the derivation arises from preserving the negative support by applying the same convolution structure over the negative real line. Specifically, if \( x_i < 0 \) for all \( i \), the estimator becomes:
\[
\widehat f(x)
= \displaystyle\frac{1}{n\,h} \sum_{i=1}^n \displaystyle\frac{1}{|x_i|}\, K\!\left( \displaystyle\frac{x}{h\,x_i} \right), \quad x < 0,
\]
where care must be taken to ensure that \( K(\cdot) \) is defined and symmetric over the appropriate range. This extension allows for consistent KDE construction on both positive and negative domains through appropriate multiplicative modeling.

\section{SHIDE}
\label{sec-SHIDE}

The symbolic models (\ref{model-symbolic}) and (\ref{model-multiplicative}) not only offer an intuitive framework for understanding and deriving the classical kernel density estimator (KDE), but also serve as foundational tools for generating pseudo data. In this section, we propose a novel density estimation method based on simulated pseudo data. Pseudo data–based approaches have appeared in the literature; for example, \cite{Cowling:1996} introduced a method to correct boundary effects in KDE by simulating data beyond the extremes of the support. In contrast, our method simulates pseudo data across the entire support of the variable, and estimates its density accordingly. The key idea is to construct a refined histogram from an expanded set of pseudo data, and then interpolate the resulting histogram using splines. We refer to this method as SHIDE, short for `Simulation and Histogram Interpolation for Density Estimation.' While conceptually straightforward, the method involves several subtle and technical challenges. 

\subsection{Pseudodata Simulation}
\label{sec-pseudodata}
To simulate pseudo data, we slightly modify Model~(\ref{model-symbolic}) as follows:
\begin{equation}
\label{model-simulation}
X'_{ij} = X_i + \varepsilon_{ij}, 
\end{equation}
for $i = 1, \ldots, n$ and $j = 1, \ldots, m$, where $m$ denotes the number of pseudo values generated for each observed data point. To proceed, we require a kernel distribution $K_h$ for the error term $\varepsilon_{ij}$. For this purpose, we introduce a family of polynomial kernels with bounded support, derived from the uniform distribution on $[-1/2, 1/2]$.
\begin{proposition}
\label{prop-polynomialKernel}
Let $U_1, \ldots, U_k$ be independent and identically distributed random variables with the uniform distribution on $[-1/2, 1/2]$, i.e., with PDF $ f_U(u) = I(-1/2 < u < 1/2). $
Define $V_k = \sum_{i=1}^k U_i, $
and let $f_k(v)$ denote the probability density function (PDF) of $V_k$. Then:
\begin{enumerate}[(i)]
    \item $f_k(v)$ is symmetric about zero, with support $[-k/2,\, k/2]$.
    \item $f_k(v)$ is a spline of degree $(k - 1)$; that is, it is a piecewise polynomial function of degree $(k - 1)$ that is continuously differentiable up to order $(k - 2)$.
    \item The characteristic function of $V_k$ is 
    \[
    \chi_k(t) = \left[\mbox{sinc}\left(\frac{t}{2}\right)\right]^k = \left[ \frac{\sin(t/2)}{t/2} \right]^k.
    \]
    with $\mbox{sinc}(x) =  \sin(x)/x.$
\end{enumerate}
\end{proposition}

\begin{figure}[h]
\centering
  \includegraphics[scale=0.30, angle=0]{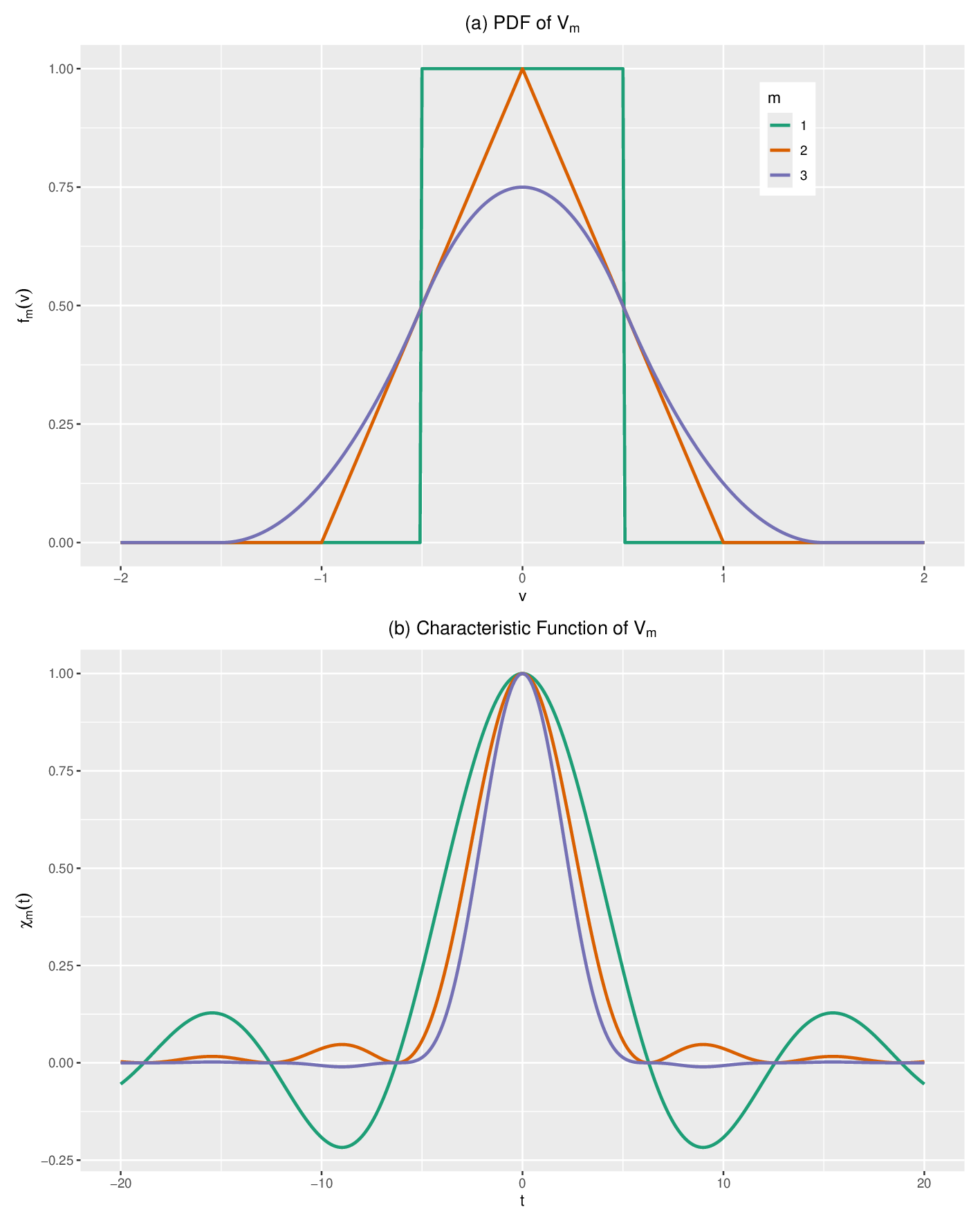}
  \caption{The PDF and Characteristic functions of $V_m$ for $m=1, 2, 3.$. \label{fig01}}
\end{figure}

From Proposition~\ref{prop-polynomialKernel}, the function $f_k(v)$ serves as a valid kernel with several desirable properties. For instance, when $k = 2$, the corresponding density is given by
\[
f_2(v) = 
\begin{cases}
1 + v, & -1 \leq v < 0, \\[6pt]
1 - v, & 0 \leq v \leq 1, \\[6pt]
0, & \text{otherwise}.
\end{cases}
\]
When $m = 3$, the density $f_3(v)$ takes the form
\[
f_3(v) = 
\begin{cases}
0, & \text{if } |v| > \dfrac{3}{2}, \\[6pt]
\dfrac{3}{4} - v^2, & -\dfrac{1}{2} \leq v < \dfrac{1}{2}, \\[6pt]
\dfrac{1}{2} \left( \dfrac{3}{2} - |v| \right)^2, & \dfrac{1}{2} \leq |v| \leq \dfrac{3}{2}.
\end{cases}
\]
Figure~\ref{fig01} displays plots of $f_k(v)$ and their corresponding characteristic functions $\chi_k(t)$ for $k = 1, 2, 3$. The $\sinc$ function plays a central role in Shannon’s sampling theorem, which establishes the theoretical foundation for reconstructing continuous signals from discrete samples. The functions $f_k(v)$ and $\chi_k(t)$, along with their mathematical properties, are well studied in approximation theory \citep{Cheney:2000}. However, to the best of our knowledge, this is the first time they are explored from a statistical viewpoint, specifically, through their interpretation as the distributions of sums of independent uniform random variables.

For our purposes, we generate the error term
$$ \varepsilon = \frac{2h}{k} V_k, $$
where \( k \in \mathbb{N} \) and \( h > 0 \) are user-specified parameters. That is, the distribution of \( \varepsilon \) corresponds to a scaled version of \( f_k \), leading to the kernel function
\begin{equation}
\label{Kh-SHIDE}
K_h(x) = \frac{2h}{k} \, f_k\left( \frac{2h x}{k} \right),
\end{equation}
with support \( [-h, h] \). Here, \( h \) serves as a smoothing parameter, playing a role analogous to the bandwidth in traditional kernel density estimation. The selection of $h$ is discussed in greater detail in Section~\ref{sec-bandwidth}.

\subsection{Histogram Interpolation with Splines}
\label{sec-interpolation}

Given the expanded dataset \( \mathcal{D}' = \{X'_{ij}\} \) of size \( n \cdot m \), we aim to estimate the underlying density using a refined histogram-based approach. Although traditional kernel density estimation (KDE) is a popular method, it often suffers from boundary bias and reduced flexibility when applied to bounded data. To address these issues, we propose an alternative method that builds on histogram interpolation.

We first construct a density histogram of \( \mathcal{D}' \), treating the \( X'_{ij} \) as independent. While observations \( \{X'_{ij} : j = 1, \ldots, m\} \) generated from the same \( X_i \) are not truly independent, they are conditionally independent given the original dataset \( \mathcal{D} \). This approach has been adopted in previous studies, such as \citet{Cowling:1996}. The histogram from \( \mathcal{D}' \) can be regarded as a refined version of that from \( \mathcal{D} \).

To choose the bin width, we use Sturges' formula \citep{Sturges:1926}, also the default in R \citep{R:2025}, $R/\{1 + 3.322 \log(n)\},$ where \( R = x_{(n)} - x_{(1)} \) is the range of the observed data $\mathcal{D}$. This formula is based on the assumption that the data distribution can be approximated by binomial coefficients, particularly when the sample size is a power of two. With the expanded dataset \( \mathcal{D}' \), the range slightly increases to \( R + 2h \), and the effective sample size becomes \( n \cdot m \). The adjusted bin width is therefore
\[
\theta = \frac{R + 2h}{1 + 3.322 \{ \log(n) + \log(m) \}},
\]
resulting in narrower bins and more intervals, especially when \( m \) is large.

We then estimate the density by interpolating the histogram based on \( \mathcal{D}' \). Let the density histogram have \( B \) bins with uniform width \( \theta \), and let \( b_1 < b_2 < \cdots < b_B \) denote the bin midpoints and \( p_1, p_2, \ldots, p_B \) the corresponding density values, satisfying \( \theta \cdot \sum_{i=1}^B p_i  = 1 \). Since a probability density function (PDF) must be nonnegative, we define \( y_i = \sqrt{p_i} \) and construct a smooth interpolating function \( S(x) \) through the points \( (b_i, y_i) \). The final density estimate is given by
\begin{equation}
\label{pdf-SHIDE}
\widetilde{f}(x) = S^2(x).
\end{equation}

We employ a natural cubic spline for interpolation \citep{Green:1994, deBoor:2001}. A cubic spline is a piecewise polynomial of degree three defined on the interval \( [b_1, b_B] \), with continuity of the function and its first two derivatives across subintervals. The natural cubic spline additionally imposes the boundary conditions \( S''(b_1) = S''(b_B) = 0 \), which help prevent spurious curvature near the boundaries. Since \( x'_{(1)} < b_1 < b_B < x'_{(n)} \), where \( x'_{(1)} \) and \( x'_{(n)} \) denote the minimum and maximum of the expanded dataset \( \mathcal{D}' \), the natural boundary condition effectively extends the smooth curve linearly beyond \( [b_1, b_B] \), covering the full data range \( [x'_{(1)}, x'_{(n)}] \).

Let \( S_i(x) \) be the cubic polynomial on the interval \( [b_i, b_{i+1}] \). Each spline segment has the form
\[
S_i(x) = c_{i0} + c_{i1}(x - b_i) + c_{i2}(x - b_i)^2 + c_{i3}(x - b_i)^3,
\]
with coefficients determined by interpolation and smoothness conditions. The uniform spacing \( \theta = b_{i+1} - b_i \) simplifies the system. Let \( M_i = S''(b_i) \) denote the second derivative at \( b_i \). Then for \( x \in [b_i, b_{i+1}] \), we can write:
\[
S_i(x) = \frac{M_{i+1}(x - b_i)^3}{6\theta} + \frac{M_i(b_{i+1} - x)^3}{6\theta} + \left( \frac{y_{i+1}}{\theta} - \frac{M_{i+1} \theta}{6} \right)(x - b_i) + \left( \frac{y_i}{\theta} - \frac{M_i \theta}{6} \right)(b_{i+1} - x).
\]
To compute \( M_i \), we solve the tridiagonal system \citep{deBoor:2001}:
\[
M_{i-1} + 4M_i + M_{i+1} = 6 \cdot \frac{y_{i+1} - 2y_i + y_{i-1}}{\theta^2}, \quad \text{for } i = 2, \ldots, B - 1,
\]
with boundary conditions \( M_1 = M_B = 0 \). Substituting the solved \( M_i \) values back into the expression for \( S_i(x) \) completes the construction of the spline. 

Natural cubic splines are especially advantageous in this setting. They produce smooth estimates with continuity up to the second derivative (\( C^2 \)), minimize curvature among twice-differentiable interpolants \citep{deBoor:2001}, and offer local control in the sense that changing a single \( p_i \) affects the spline only locally. Moreover, the natural boundary conditions reduce edge oscillations and improve behavior at the distribution tails \citep{Green:1994}.

\subsection{Extension to Positive/Bounded Variables}
\label{sec-bounded}
One notable advantage of SHIDE over traditional kernel density estimation (KDE) is its greater flexibility in handling specialized random variables, such as those that are positive, negative, or bounded. These characteristics can be naturally addressed during the simulation stage through appropriate transformations.

If the variable \( X \) is bounded below by \( L \), with the special case \( L = 0 \) corresponding to positive-valued data, we apply a logarithmic transformation:
\[
X := \log(X - L) \in \mathbb{R}.
\]
Error terms \( \varepsilon \) are then generated based on the transformed variable. After computing the perturbed value \( X' := X + \varepsilon \), we transform back to the original scale:
\[
X' := \exp(X') + L.
\]

If \( X < U \) is bounded above, we instead consider \( -X > -U \) and proceed analogously by applying the logarithmic transformation to \( -X \), followed by back-transformation.

For variables bounded on both sides, \( L < X < U \), we first normalize the range to \( (0,1) \) using $X := (X - L)/(U - L),$ and then apply the logit transformation:
\[
X := \log\left( \frac{X}{1 - X} \right) \in \mathbb{R}.
\]
After adding error \( \varepsilon \) to the transformed variable, i.e., \( X' := X + \varepsilon \), we transform back to the original scale via the expit function:
\[
X' := \frac{\exp(X')}{1 + \exp(X')} \cdot (U - L) + L.
\]

Histogram interpolation is then applied to the expanded dataset \( \mathcal{D}' = \{X'_{ij}\} \). Note that an alternative strategy for handling positive or negative variables is to use the multiplicative model~(\ref{model-multiplicative}) and draw error terms from positive distributions, such as the gamma or log-normal distributions. However, the transformation-based approach outlined above, grounded in the additive model~(\ref{model-symbolic}), offers a unified framework for treating both one-sided and two-sided bounded variables.

\section{Theoretical Analysis}
\label{sec-theory}

This section provides a rigorous theoretical analysis of SHIDE. We establish asymptotic properties including pointwise and uniform consistency, bias-variance decomposition, and mean integrated squared error (MISE). Comparisons with classical KDE demonstrate SHIDE's advantages in boundary adaptation and computational efficiency.

Let $X_1,\dots,X_n \stackrel{\text{i.i.d.}}{\sim} f$ with density $f$ supported on $\mathcal{S}\subseteq\mathbb{R}$. For a fixed integer $k\ge 1$, let $U_1,\dots,U_k \stackrel{\text{i.i.d.}}{\sim}\mathrm{Unif}[-1/2, 1/2]$, set $V_k=\sum_{\ell=1}^k U_\ell$ with density $f_k$, and generate $m$ pseudo–observations per datum via
\[
X'_{ij}=X_i+\varepsilon_{ij},\qquad \varepsilon_{ij}=\frac{2h}{k}\,V_k,\quad i=1,\dots,n,\;j=1,\dots,m,
\]
so that the error density is the compactly supported kernel
\[
K_h(x)=\frac{k}{2h}\,f_k\!\left(\frac{kx}{2h}\right),\qquad \mathrm{supp}(K_h)=[-h,h].
\]
Thus $K_h$ is the rescaled $f_k$, correctly normalized under the change of variables $x=(2h/k)\,v$.

\begin{assumption}\label{assump:f}
The true density $f$ satisfies $f\in C^2(\mathbb{R})$ and is bounded and bounded away from zero on compact subsets of $\mathcal{S}$.
\end{assumption}

\begin{assumption}\label{assump:kernel}
The kernel $K_h$ satisfies: (i) $\operatorname{supp}(K_h)=[-h,h]$; (ii) $\int K_h(x)\,dx=1$; (iii) $\int x\,K_h(x)\,dx=0$; (iv) $\int x^2 K_h(x)\,dx=\sigma_K^2\,h^2<\infty$, where $\sigma_K^2\in(0,\infty)$ is the (scaled) second moment.
\end{assumption}

Clearly, the SHIDE kernel $K_h(\cdot)$ satisfies Assumption~\ref{assump:kernel}. Indeed, since $V_k=\sum_{i=1}^k U_i$ has $\operatorname{Var}(V_k)=k/12$ and $\varepsilon=(2h/k)\,V_k$, we obtain
\[
\int x^2 K_h(x)\,dx=\operatorname{Var}(\varepsilon)=\frac{4h^2}{k^2}\cdot\frac{k}{12}=\frac{h^2}{3k},
\]
so $\sigma_K^2=1/(3k)$. We nevertheless retain Assumption~\ref{assump:kernel} to state results for any compactly supported, symmetric kernel with finite second moment, thereby extending the analysis beyond this specific construction.

Next, a histogram is constructed for the pseudo–sample $\mathcal{D}'=\{X'_{ij}\}$ with uniform bin width $\theta>0$ and bin midpoints $b_1<\cdots<b_B$. Let
$$
N_r = \sum_{i=1}^n \sum_{j=1}^m I\{X'_{ij}\in \text{bin } r\}
\mbox{~~~and~~~}
p_r=\frac{N_r}{nm\,\theta}, \mbox{~~for~} r=1,\dots,B,
$$
so that $\sum_{r=1}^B p_r=1$. Define $y_r=\sqrt{p_r}$ and let $S$ be the natural cubic spline interpolant of the points $\{(b_r,y_r)\}_{r=1}^B$. The SHIDE estimator is then
\begin{equation}\label{eq:pdf-SHIDE}
\widetilde{f}(x)=[S(x)]^2,\qquad x\in[b_1,b_B],
\end{equation}
with the natural boundary conditions $S''(b_1)=S''(b_B)=0$ implicitly yielding a linear continuation outside $[b_1,b_B]$.

\subsection{Asymptotic Properties}
\label{subsec-AsymProp}

We first establish pointwise consistency of the SHIDE estimator. 

\begin{theorem}\label{thm:pointwise}
Let Assumptions~\ref{assump:f}–\ref{assump:kernel} hold and let $x\in\mathrm{int}(\mathcal{S})$. Suppose $h\to0$, $\theta\to0$, $m\to\infty$, and $n\to\infty$ with $nm\,\theta\to\infty$. Then
\[
\widetilde{f}(x)\;\xrightarrow{p}\; f(x).
\]
\end{theorem}
We defer all proofs to the Supplement. The key step is the decomposition
\[
\bigl|\widetilde{f}(x)-f(x)\bigr|
\;\le\;
\bigl|\widetilde{f}(x)-f_h(x)\bigr|
\;+\;
\bigl|f_h(x)-f(x)\bigr|,
\]
where \(f_h=f*K_h\) is the convolution of \(f\) with the kernel \(K_h\). This separates the deterministic convolution bias from the stochastic error arising from the refined histogram and the spline interpolation. The second term is a deterministic bias of order \(h^2\) (by symmetry and the finite second moment of \(K_h\)), while the first term converges to zero in probability because the refined histogram consistently estimates \(f_h\) at bin midpoints when \(\theta\to 0\) and \(nm\,\theta\to\infty\), and the natural cubic spline interpolant approximates \(C^2\) functions on a uniform grid with error \(O(\theta^2)\); in addition, the nodal (binwise) stochastic errors are \(o_p(1)\) by histogram consistency. Pointwise consistency validates SHIDE as a statistically sound estimator at interior points of the support. It also clarifies the roles of the tuning parameters: $h$ controls the convolution bias through the pseudo–data kernel, while $\theta$ governs discretization and interpolation error; the requirement $nm\,\theta\to\infty$ guarantees variance reduction from the enlarged sample. 

We next move on to a pointwise bias–variance analysis of the SHIDE estimator. The goal is to separate the deterministic bias terms—arising from the pseudo–data convolution and from binning/interpolation—from the stochastic variance terms due to finite pseudo–sample size and histogram discretization.

\begin{theorem}\label{thm:biasvar}
Under the same assumptions as Theorem~\ref{thm:pointwise}, we have
\[
\mathbb{E}\big[\widetilde{f}(x)\big]-f(x)
\;=\;
\underbrace{\frac{1}{2}\,\sigma_K^2 h^2 f''(x)}_{\text{convolution bias}}
\;+\;
\underbrace{O(\theta^2)}_{\text{binning/interpolation bias}}
\;+\;
o\!\big(h^2+\theta^2\big),
\]
and
\[
\mathrm{Var}\big(\widetilde{f}(x)\big)
\;=\;
\underbrace{\frac{f(x)}{nm\,\theta}\int K_h^2(u)\,du}_{\text{histogram (within–sample) variance}}
\;+\;
\underbrace{\frac{1}{n}\,\mathrm{Var}\!\big(K_h(x-X_1)\big)}_{\text{between–sample variance}}
\;+\;
o\!\left(\frac{1}{nm\,\theta}+\frac{1}{n}\right).
\]
\end{theorem}

The bias decomposition isolates two sources. The term $(1/2) \,\sigma_K^2 h^2 f''(x)$ is the usual second–order convolution bias from smoothing the data with a symmetric, finite–variance kernel; it vanishes as $h\to 0$. The $O(\theta^2)$ term arises from midpoint discretization and the spline interpolation error on a uniform grid for $C^2$ targets. On the variance side, the leading term $(nm\,\theta)^{-1}\int K_h^2$ reflects binwise counting variability in the enlarged pseudo–sample, while the $n^{-1}$ term captures variability across i.i.d.\ draws of the original sample through the smoothed summand $K_h(x-X_i)$.

These formulas clarify tuning in SHIDE. The parameter $h$ controls convolution bias; the bin width $\theta$ controls discretization and interpolation error and, jointly with $nm$, dictates within–sample variance. Balancing $h^2$ against $(nm\,\theta)^{-1}$ and $\theta^2$ underlies the MISE–optimal choices developed next and explains why SHIDE attains the same $n^{-4/5}$ rate as second–order kernel estimators under standard smoothness conditions.


Now we consider the mean integrated squared error (MISE) of the SHIDE estimator. The analysis combines the above pointwise bias–variance expansion with integration over the domain, and makes explicit how the tuning parameters $(h,\theta)$ and the pseudo–sample size $nm$ govern the dominant terms.

\begin{theorem}\label{thm:mise}
Under the conditions of Theorem~\ref{thm:pointwise}, it follows that
\begin{eqnarray*}
\mathrm{MISE}(\widetilde{f})
&:=& \mathbb{E}\!\int\!\bigl(\widetilde{f}(x)-f(x)\bigr)^2\,dx \\[6pt]
&=& \underbrace{\frac{1}{4}\sigma_K^4 h^4 \int (f''(x))^2\,dx}_{\text{integrated convolution bias}^2}
\;+\;
\underbrace{\dfrac{1}{nm\,\theta}\int K_h^2(u)\,du}_{\text{integrated within–sample variance}} \\[6pt]
&& \;+\;
\underbrace{\dfrac{1}{n}\int \Bigl\{\mathbb{E}\bigl[K_h(x-X_1)^2\bigr]
- \bigl(\mathbb{E}[K_h(x-X_1)]\bigr)^2\Bigr\}\,dx}_{\text{between–sample variance}}
\;+\; R_n.
\end{eqnarray*}
where
\[
\int K_h^2(u)\,du=\frac{1}{h}\int K^2(t)\,dt,\qquad
\int \bigl(\mathbb{E}[K_h(x-X_1)]\bigr)^2 dx=\|f*K_h\|_2^2,
\]
and the remainder satisfies
\[
R_n=o\!\big(h^4\big)+o\!\Big(\frac{1}{nm\,\theta}\Big)+o\!\Big(\frac{1}{n}\Big).
\]
Equivalently,
\begin{equation}
\label{MISE}
\mathrm{MISE}(\widetilde{f})
=\frac{1}{4}\sigma_K^4 h^4 \int (f'')^2
+\frac{R(K)}{nm\,h\,\theta}
+\frac{1}{n}\Bigl(\frac{R(K)}{h}-\|f*K_h\|_2^2\Bigr)
+R_n,
\end{equation}
with $R(K)=\int K^2$ and $\|\,\cdot\,\|_2$ the $L^2$ norm.
\end{theorem}

The expression separates three leading contributions. The term $(1/4) \sigma_K^4 h^4\int(f'')^2$ is the integrated squared convolution bias and decays as $h\to0$. The term $(nm\,\theta)^{-1}\int K_h^2$ is the integrated within–sample variance arising from binwise counting noise in the enlarged pseudo–sample; it improves with larger $m$ and finer bin width $\theta$, but scales as $h^{-1}$ through $\int K_h^2$. The final term, $(1/n)\{(1/h)R(K)-\|f*K_h\|_2^2\}$, reflects between–sample variability due to the original sample; its leading order is $(nh)^{-1}$.

These formulas guide tuning. Choosing $\theta$ of order $h$ keeps the binning/interpolation bias commensurate with the convolution bias, and balancing $h^4$ against $(nh)^{-1}$ recovers the familiar $h\asymp n^{-1/5}$ rate; with $m$ fixed or slowly increasing, the within–sample term contributes $(nm\,h\,\theta)^{-1}$ and recommends $\theta\asymp h$ to match orders. Under these choices, $\mathrm{MISE}(\widetilde{f})=O(n^{-4/5})$, matching the optimal rate for second–order kernel estimators \citep[see, e.g.,][]{Tsybakov:2009} while retaining SHIDE’s advantages for constrained supports and modality preservation established elsewhere in the analysis.

\subsection{Comparison with KDE}
\label{subsec-comparison}

The asymptotic properties of SHIDE closely parallel those of the traditional KDE \citep[see, e.g.,][]{Silverman:1986, Wand:1995}. Both methods achieve pointwise consistency under mild smoothness conditions, and both exhibit the classical bias–variance tradeoff: a squared bias term of order $h^4$ and a variance term of order $(nh)^{-1}$ when tuning parameters are chosen optimally. Consequently, SHIDE attains the minimax rate of convergence for second–order kernels, with $\mathrm{MISE} = O(n^{-4/5})$, identical to KDE.

Despite these similarities, important differences remain. Traditional KDE suffers from boundary bias on constrained supports and requires special corrections (reflection, boundary kernels, etc.). In contrast, SHIDE incorporates compactly supported kernels and transformation-based pseudo–data generation, which naturally adapt to bounded or semi-bounded domains. Moreover, the histogram–spline interpolation in SHIDE introduces an additional discretization parameter $\theta$, but under the scaling $\theta \asymp h$ this does not degrade the convergence rate and instead affords greater flexibility in controlling local smoothness. In practice, these features enable SHIDE to better capture multi-modal structures and to preserve distributional features in regions where KDE may oversmooth or distort the density near boundaries \citep{Schuster:1985, Chen:1999}.


We consider a formal comparison at the boundary of the support. Throughout, assume $\mathcal{S}=[a,\infty)$ for notational clarity; the upper–boundary case is analogous. Define $F_K(c):=\int_{-1}^{c} K(t)\,dt$ and $\mu_1^-(K;c):=\int_{-1}^{c} t\,K(t)\,dt$ for a symmetric, compactly supported base kernel $K$ with support $[-1,1]$ and zero first moment.

\begin{proposition}\label{prop:boundary}
Let $x=a+c\,h$ with fixed $c\in[0,1)$ (a point within $O(h)$ of the boundary). Under Assumptions~\ref{assump:f}–\ref{assump:kernel} and the support–preserving pseudo–data construction (via transformations when needed), the SHIDE estimator satisfies
\[
\bigl|\widetilde{f}(x)-f(x)\bigr| \;=\; O_p\!\big(h^2+\theta^2\big),
\]
whereas the uncorrected symmetric–kernel KDE obeys
\[
\mathbb{E}\!\big[\widehat{f}_{\mathrm{KDE}}(x)\big]-f(x)
= \big(F_K(c)-1\big)\,f(x)\;-\;h\,f'(x)\,\mu_1^-(K;c)\;+\;O(h^2),
\]
so that, in general, the leading boundary bias is of order $1$ (via the mass–deficit term $F_K(c)-1\neq 0$ for $c<1$).
\end{proposition}

The proof in Section~\ref{proof:boundary} (Supplement) highlights two distinct effects. For SHIDE, the only systematic boundary bias is the $O(h^2)$ convolution term induced by the compact, symmetric kernel; the histogram–spline step adds at most $O(\theta^2)$, thanks to the natural boundary condition $S''(a)=0$ on a uniform grid. By contrast, uncorrected KDE loses kernel mass beyond the support, yielding a nonvanishing term $\big[F_K(c)-1\big]\,f(x)$ and thus an $O(1)$ boundary bias for points $x$ within $O(h)$ of the boundary.

This comparison clarifies SHIDE’s boundary adaptation: after a support–preserving pseudo–data step, the spline interpolation with natural boundary conditions eliminates the dominant boundary distortion that afflicts KDE. Consequently, SHIDE attains the same interior order as in the bulk, $O_p(h^2+\theta^2)$, right up to the boundary, whereas KDE requires explicit boundary corrections (e.g., reflection or boundary kernels) to avoid an $O(1)$ loss.

We next move on to a careful comparison of computational cost. Let $n$ be the sample size, $m$ the number of pseudo–replicates per observation, $k$ the order of the uniform–sum (polynomial) kernel used to generate errors, $B$ the number of histogram bins, and $G$ the number of evaluation points at which the final density is returned.

\begin{proposition}\label{prop:complexity}
Under a uniform bin grid and natural cubic spline interpolation, the arithmetic cost of SHIDE decomposes as
\begin{eqnarray}
\mathrm{Cost}(\text{SHIDE})
&=& \underbrace{O(n m k)}_{\text{pseudo–data generation}}
+ \underbrace{O(n m)}_{\text{single–pass binning}}  + \underbrace{O(B)}_{\text{spline setup}}
+ \underbrace{O(B)}_{\text{spline solve}}
+ \underbrace{O(G)}_{\text{evaluation}} \nonumber \\[6pt]
&=& O(n m k + n m + B + G). \nonumber
\end{eqnarray}
In particular, for fixed $k=O(1)$ and $m=O(1)$ with $B=O(\log n)$ (e.g., Sturges’ rule), one has $\mathrm{Cost}(\text{SHIDE})=O(n+G)$.

Direct KDE \citep{Silverman:1986, Wand:1995} evaluated on $G$ points has cost
\[
\mathrm{Cost}(\text{KDE}_{\mathrm{direct}})=O(nG),
\]
while binned/FFT KDE \citep{Silverman:1982} on a regular grid satisfies
\[
\mathrm{Cost}(\text{KDE}_{\mathrm{FFT}})=O(n+G\log G).
\]
Consequently, with $m=O(1)$ and $B=O(\log n)$, SHIDE is asymptotically faster than direct KDE whenever $G\gtrsim \log n$, and it is comparable to FFT–KDE in the regime where a regular grid is used for evaluation.
\end{proposition}

Two clarifications help interpret these bounds. First, the spline stage in SHIDE is \emph{linear} in $B$ (tridiagonal), not $O(B^3)$; there is no cubic bottleneck for natural cubic spline interpolation on a line. Second, whether SHIDE is actually faster depends on the evaluation regime and parameter scaling. If a density is returned on a moderately fine grid ($G\gg 1$), direct KDE incurs $O(nG)$ work, while SHIDE remains $O(n)$ in $n$ for fixed $m,k$ and slowly growing $B$ (e.g., $O(\log n)$), plus an $O(G)$ evaluation term. Compared to FFT–KDE, SHIDE replaces the $G\log G$ grid convolution by $O(B+G)$; the preferable method can depend on constants and grid resolution.

A few caveats ensure the claims hold uniformly. If $G$ is very small (e.g., a constant number of query points), direct KDE can be cheaper than fitting a histogram and spline, so the advantage of SHIDE is not universal at tiny $G$. If a nonuniform bin grid is used, locating bins may cost $O(\log B)$ per query unless a direct map is precomputed, making evaluation $O(G\log B)$ (still mild when $B=O(\log n)$). Finally, if $m$ grows with $n$ (e.g., $m=n^\alpha$), the simulation and binning pass becomes $O(n^{1+\alpha})$; the “faster than direct KDE’’ conclusion should then be restricted to $m=O(1)$ (or at least $m=o(G)$) to preserve the stated advantage.

\section{Bandwidth Selection}
\label{sec-bandwidth}

Bandwidth selection represents a critical aspect of kernel density estimation (KDE), as it directly governs the bias-variance trade-off in the resulting density estimate. The foundational work of \citet{Rosenblatt:1956} and \citet{Parzen:1962} established the KDE framework, yet the challenge of selecting an optimal bandwidth has evolved into a major research domain. Early developments included rule-of-thumb methods based on AMISE, most notably the plug-in and normal reference rules proposed by \citet{Silverman:1986}. Subsequent research introduced data-driven alternatives through cross-validation techniques, including least-squares and biased cross-validation, which adapt to unknown data structures \citep{Rudemo:1982, Bowman:1984}. More sophisticated plug-in approaches that estimate the unknown density curvature were advanced by \citet{Sheather:1991} and have gained general recommendation in the literature \citep{Venables:2002}. Later developments explored adaptive and local bandwidth strategies, enhancing estimation performance in regions of varying data density \citep{Terrell:1992}. Despite these methodological advances, bandwidth selection remains one of the most influential and nuanced components of KDE.

In the SHIDE framework, the use of a compact-support kernel density for generating the error term $\varepsilon$ is motivated by the need to preserve the underlying distributional structure of the original data. The half-width $h$ of $\varepsilon$'s support functions as a bandwidth parameter, providing direct control over the spread of simulated data around each observed value. Building upon the asymptotic properties established in Section~\ref{sec-theory}, we develop theoretical guidance for bandwidth selection in SHIDE. We investigate two complementary approaches: (i) an optimal bandwidth derived by minimizing the asymptotic mean integrated squared error (AMISE), and (ii) a percentile-based rule grounded in empirical nearest-neighbor spacing.

\subsection{AMISE-Optimal Selector}
\label{subsec-bandwidth-opt}

We begin by deriving the optimal pseudo-data bandwidth \(h\) via minimization of the AMISE, the leading-order approximation to the MISE. This derivation clarifies how the pseudo-sample size \(m\) and the histogram bin width \(\theta\) jointly govern the bias-variance trade-off.

Define
\[
\Psi(f'') := \int (f''(x))^2\,dx,
\qquad
R(K) := \int K^2(u)\,du,
\]
and recall that $\int K_h^2(u)\,du = h^{-1}R(K)$ and 
$\sigma_K^2 = \int u^2 K(u)\,du$. 
From Theorem~\ref{thm:mise}, the dominant AMISE terms in (\ref{MISE}) are
\[
\mathrm{AMISE}(h,\theta,m)
= \frac{1}{4}\,\sigma_K^4\,h^4\,\Psi(f'')
+ \frac{R(K)}{n h}
+ \frac{R(K)}{n m \theta\, h}
+ o\!\left(h^4+\frac{1}{n h}+\frac{1}{n m \theta\, h}\right).
\]
Under the coupling $m\theta \to c\in(0,\infty)$ (constant pseudo--observations per effective bin),
\[
\mathrm{AMISE}(h)
= \frac{1}{4}\,\sigma_K^4\,h^4\,\Psi(f'')
+ \frac{R(K)}{n h}\Bigl(1+\frac{1}{c}\Bigr)
+ o\!\Bigl(h^4+\frac{1}{n h}\Bigr).
\]

\begin{theorem}
\label{thm:optimalh}
Assume $f\in C^2$ with $\Psi(f'')<\infty$, and that Assumption~\ref{assump:kernel} holds. 
In the regime $h\to0$, $\theta\to0$, $n\to\infty$, $m\to\infty$, $nm\theta\to\infty$, with $m\theta\to c\in(0,\infty)$, 
the AMISE--optimal bandwidth is
\begin{equation}
\label{h-opt}
h_{\mathrm{opt}}
= \left(\frac{R(K)\left(1+1/c \right)}{\sigma_K^4\,\Psi(f'')}\right)^{1/5}\,n^{-1/5}\,(1+o(1)),
\end{equation}
and the corresponding minimum is
\begin{equation}
\label{AMISE-opt}
\mathrm{AMISE}\bigl(h_{\mathrm{opt}}\bigr)
= \frac{5}{4}\,
\bigl(\sigma_K^4\,\Psi(f'')\bigr)^{1/5}\,
\bigl(R(K)\bigr)^{4/5}\,
\left(1+\frac{1}{c}\right)^{4/5}\,
n^{-4/5}\,(1+o(1)).
\end{equation}
Thus SHIDE attains the classical $n^{-4/5}$ rate for second--order kernels; the factor $(1+1/c)^{4/5}$ 
reflects the pseudo--sample--per--bin coupling.
\end{theorem}

These formulas mirror the classical AMISE results for second--order kernels, with a transparent modification due to pseudo--data and binning. The coupling $m\theta\to c$ ensures that the within--sample counting term matches the 
between--sample variance in order, so the usual balance $h^4$ versus $(nh)^{-1}$ yields $h_{\mathrm{opt}}\asymp n^{-1/5}$ and $\mathrm{AMISE}\asymp n^{-4/5}$. 
Practically, one may enforce $\theta\asymp h$ and $m\asymp 1/h$ so that $m\theta\approx c\in[0.5,2]$.  If $m$ is fixed while $\theta\asymp h$, then $m\theta\to0$ and the within--sample term dominates as $(n m h^2)^{-1}$, 
leading to a slower rate; hence a mild growth of $m$ (or a coarser $\theta$) is essential for SHIDE to retain the optimal $n^{-4/5}$ behavior.

\subsection{Selector Based on Nearest-Neighbor Spacing}
\label{subsec-bandwidth-per}

In SHIDE, the half-width $h$ of the support of $\varepsilon$ directly controls the spread of pseudo-samples around each observation. 
If $h$ is too large, outliers may be blurred by simulated values, diminishing their influence. 
A practical and intuitive choice for $h$ is based on the first-order spacings of the ordered data. Let $x_{(1)} \le \cdots \le x_{(n)}$ be the order statistics and define $d_i = x_{(i+1)} - x_{(i)}$. 
Setting $$h = d_\alpha/2,$$ where $d_\alpha$ is the empirical $\alpha$-quantile of $\{d_i\}$, ensures that no more than an $\alpha\%$ fraction of pseudo-samples overlap, maintaining separation among well-spaced points.

\begin{figure}[h]
\centering
  \includegraphics[scale=0.40, angle=0]{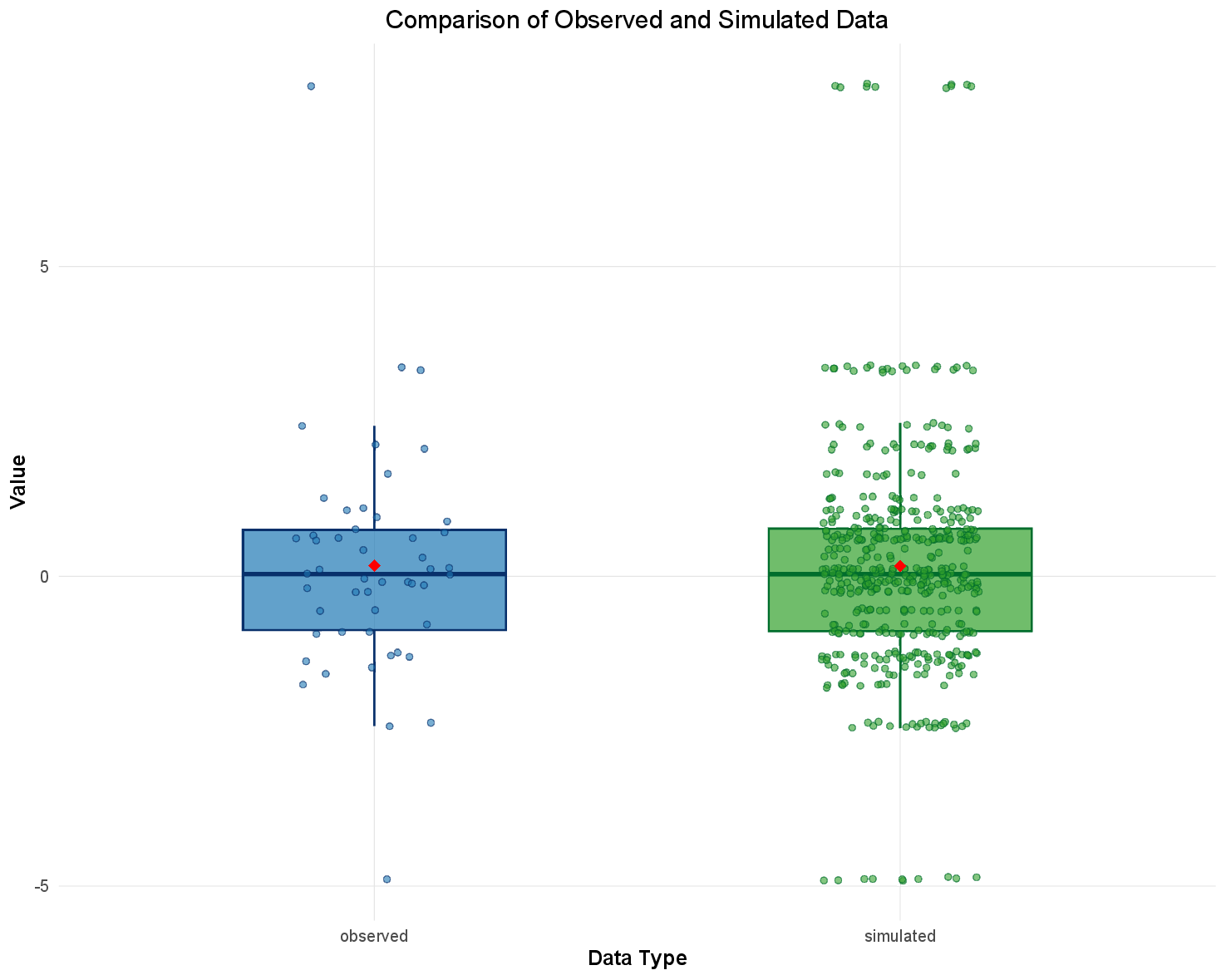}
  \caption{Illustration of Observed and Simulated Data. The observed data consist of $n = 50$ values generated from a $t(3)$ distribution, exhibiting two apparent outliers. The simulated data comprise 500 values, created by adding random noise to each observed value. 
\label{fig02}}
\end{figure}

Figure~\ref{fig02} illustrates this idea using $n=50$ observations from a $t(3)$ distribution containing two visible outliers. 
With $k=3$ and $\alpha=0.50$, the simulated data show that while moderate overlap occurs in the central cluster, the outliers remain clearly separated from the main body.

We now formalize the large-sample behavior of this spacing-based rule. 
Specifically, we study the asymptotic expansion of the spacing quantile $d_\alpha$ to show that while the rule yields consistent estimators, it is asymptotically undersmoothed relative to the AMISE-optimal bandwidth.

\begin{proposition} 
\label{prop:percentile-raw}
Suppose Assumption~\ref{assump:f} holds with $f$ continuous and strictly positive on compact subintervals of $\mathrm{int}(\mathcal S)$. 
Let $d_\alpha$ be the empirical $\alpha$-quantile of $\{d_i\}_{i=1}^{n-1}$ for a fixed $\alpha\in(0,1)$, and set $h_n:=d_\alpha/2$. Then
\[
d_\alpha \;=\; \frac{q_\alpha}{n\,\bar f_\alpha}\,\{1+o_p(1)\}
\mbox{~~~with~~}
q_\alpha:=-\log(1-\alpha),
\]
for some random $\bar f_\alpha$ satisfying
$\bar f_\alpha\in\bigl[\inf_{x\in\mathcal I_\alpha} f(x),\,\sup_{x\in\mathcal I_\alpha} f(x)\bigr]$
with probability tending to one, where $\mathcal I_\alpha\subset\mathrm{int}(\mathcal S)$ is an interior interval determined by the empirical location of the $\alpha$-spacing among $\{d_i\}$. In particular, $h_n\to0$ and $n h_n\to0$, so SHIDE with $h=h_n$ is pointwise consistent but undersmooths relative to the $n^{-1/5}$ AMISE-optimal rate.
\end{proposition}

Note that the factor $q_\alpha=-\log(1-\alpha)$ arises from the exponential limit for scaled uniform spacings: for interior indices $i=\lfloor n\tau\rfloor$, $n\,\Delta_i\xrightarrow{d}\mathrm{Exp}(1)$, whose $\alpha$-quantile is $q_\alpha$; see \cite{David:2003}. Proposition~\ref{prop:percentile-raw} shows that the unscaled rule $h=d_\alpha/2$ picks a bandwidth of order $1/n$. This guarantees consistency but is smaller than the AMISE-optimal order $n^{-1/5}$ and therefore leads to undersmoothing.

To recover optimal rates, we next turn to a calibrated spacing rule that preserves the practical appeal of percentile tuning while attaining AMISE-optimal performance. The idea is to rescale the raw spacing quantile by an $n$-dependent factor and mild pilot constants so that the resulting bandwidth matches the AMISE-optimal target in Theorem~\ref{thm:optimalh}.

Define
\begin{equation}
\label{h-calib}
h_n^{\mathrm{perc}}
:= \lambda_n\,d_\alpha 
\mbox{~~~with~}
\lambda_n
:= n^{4/5}\,
\left(\frac{R(K)\,(1+1/c)}{\sigma_K^4\,\widehat\Psi}\right)^{1/5}\,
\frac{\widehat f(x_\alpha)}{q_\alpha},
\end{equation}
where $d_\alpha$ is the empirical $\alpha$–quantile of the nearest–neighbor spacings $\{d_i\}_{i=1}^{n-1}$, $q_\alpha=-\log(1-\alpha)$, $\widehat f(x_\alpha)$ is a pilot estimator of $f$ at a representative interior location $x_\alpha$, and $\widehat\Psi$ is a pilot for $\Psi(f'')=\int (f''(x))^2\,dx$. The factor $n^{4/5}$ bridges the $1/n$ scale of spacings and the $n^{-1/5}$ optimal bandwidth; the pilot terms align the constant with AMISE.

\begin{proposition}
\label{prop:percentile-calibrated}
Assume $f\in C^2$ with $\Psi(f'')<\infty$, the kernel satisfies Assumption~\ref{assump:kernel}, and the coupling $m\theta\to c\in(0,\infty)$ with $nm\theta\to\infty$ holds. Suppose $\widehat f(x_\alpha)\xrightarrow{p} f(x_\alpha)$ and $\widehat\Psi\xrightarrow{p}\Psi(f'')$, with $x_\alpha$ chosen in an interior region where $f$ is continuous and bounded away from zero. Then
\[
\frac{h_n^{\mathrm{perc}}}{\,h_{\mathrm{opt}}\,}\;\overset{p}{\longrightarrow}\;1,
\]
where $h_{\mathrm{opt}}=\Bigl(\frac{R(K)(1+1/c)}{\sigma_K^4\,\Psi(f'')}\Bigr)^{1/5} n^{-1/5}$ is the AMISE–optimal bandwidth from Theorem~\ref{thm:optimalh}. Consequently,
\[
\mathrm{AMISE}\bigl(h_n^{\mathrm{perc}}\bigr)
=\mathrm{AMISE}\bigl(h_{\mathrm{opt}}\bigr)\,\{1+o(1)\}
=O(n^{-4/5}).
\]
\end{proposition}

The construction scales the raw spacing by $n^{4/5}$ to convert the intrinsic $1/n$ spacing magnitude into the $n^{-1/5}$ bandwidth scale, and then inserts light pilot corrections to match the AMISE constant. The proof hinges on the exponential limit for uniform spacings, the mean-value relation between spacings on the $X$- and $U=F(X)$-scales, and Slutsky’s theorem. Practically, one may take $\alpha$ in a central range (for example $\alpha\in[0.2,0.5]$), choose $x_\alpha$ as the sample median or a central quantile, and estimate $\Psi(f'')$ from the SHIDE spline. Under these mild choices, the calibrated percentile selector is asymptotically equivalent to the AMISE-optimal bandwidth while retaining the interpretability and robustness of spacing-based tuning.

\section{Numerical Results}
\label{sec-numerical}

This section discusses key implementation issues for SHIDE, followed by simulation studies evaluating its performance relative to KDE.

\subsection{Implementation Issues}
\label{subsec-implemenation}

Implementing SHIDE requires estimating the two theoretical bandwidths from Section~\ref{sec-bandwidth}: 
(i) the AMISE--optimal bandwidth $h_{\mathrm{opt}}$ in (\ref{h-opt}), and 
(ii) the calibrated percentile-based bandwidth $h_n^{\mathrm{perc}}$ in (\ref{h-calib}). 
Both involve estimating unknown quantities such as $\Psi(f'')$ and the density $f(x)$ at selected points.

With the kernel $K_h(x)$ defined in (\ref{Kh-SHIDE}) of Section~\ref{sec-pseudodata}, it can be shown that
$$
R(K) = \frac{k}{2 \cdot 4^k}\, \binom{2k}{k}
\qquad \text{and} \qquad
\sigma_K^2 = \frac{1}{3k}.
$$
For practical use, setting $\theta \approx h$ and $m \approx 1/h$ implies $m\theta \approx 1$, meaning approximately one pseudo-observation per histogram bin. This yields $c \approx 1$, which simplifies the theoretical expressions and is used as the default coupling. 
In practice, any $c \in [0.5, 2]$ provides stable results.

To estimate $\Psi(f'') = \int (f''(x))^2\,dx$, a smooth pilot density estimate is required. A simple option is to use a rule-of-thumb bandwidth such as Silverman’s rule, or an oversmoothed SHIDE estimate, to obtain a pilot $\widehat f(x)$ from which $f''(x)$ can be evaluated numerically on a fine grid. Because $\Psi(f'')$ is relatively insensitive to the exact pilot choice, rough defaults perform well.  Alternatively, assuming normality provides a fast approximation: for a normal density with standard deviation $s$, $\Psi(f'') = 3s^{-5}/(8\sqrt{\pi})$. Thus, one may set $s = \mathrm{sd}(x)$ or use a robust scale estimate $s = \mathrm{IQR}(x)/1.349$ and compute $\Psi(f'')$ directly, following the normal-reference approach of \citet{Silverman:1986}.

These components allow estimation of $h_{\mathrm{opt}}$. 
To compute $h_n^{\mathrm{perc}}$, a convenient default is $\alpha = 0.5$ (the median spacing), though any $\alpha \in [0.2, 0.5]$ is also acceptable. An open-source implementation of SHIDE in \textsf{R} \citep{R:2025} is available at
\newline\centerline{\url{https://github.com/xgsu/SHIDE},}
where all the options discussed above are supported.

\subsection{Simulation Studies}
\label{subsec-simulation}

To assess the performance of SHIDE and compare it with KDE, we generated data $x$ from five representative models, as tabulated below. These models encompass a range of scenarios: (I) a standard unimodal case, (II) a bimodal distribution, (III) a heavy-tailed model with outliers, (IV) a positive-support distribution, and (V) a bounded-support setting.
\begin{center}
\begin{tabular}{cll}
\toprule
Model & Description & Feature \\ 
\midrule
I & $\mathcal{N}(0,1)$ & normal, unimodal \\
II & $0.35\,\mathcal{N}(-1,1) + 0.65\,\mathcal{N}(2,2)$ & bimodal \\
III & Cauchy$(0,1)$ & heavy-tailed \\
IV & Exponential$(\lambda=1)$ & positive support \\
V & Truncated $\mathcal{N}(0,3)$ on $(a=-1,b=0.5)$ & bounded \\
\bottomrule
\end{tabular}
\end{center}

Two sample sizes, $n \in \{50, 500\}$, were considered. For each generated dataset, both KDE and SHIDE were applied to estimate the density function. 
KDE used the widely recommended Sheather--Jones (SJ) bandwidth selector \citep{Sheather:1991}, while SHIDE employed both $h_{\mathrm{opt}}$ and $h_n^{\mathrm{perc}}$ as described in Section~\ref{sec-bandwidth}. 
For SHIDE, the pseudo-data generation used $m=10$ replications per observation with default coupling constant $c=1$. 
Each model configuration was replicated 300 times.

\renewcommand{\tabcolsep}{8.0pt}
\renewcommand{\arraystretch}{1.1}
\renewcommand{\baselinestretch}{1.0}
\begin{table}[h]
\caption{Comparison of MISE between SHIDE and KDE methods. Median and MAD (median absolute deviation) are computed over 300 simulation runs. SHIDE uses two bandwidth selectors: AMISE-optimal (opt) and nearest-neighbor spacing percentile (per). KDE uses the SJ plug-in selector \citep{Sheather:1991}. \label{tbl01-MISE}}
\centering
\begin{tabular}[t]{clcrrcrr}
\toprule
& & Bandwidth & \multicolumn{2}{c}{$n=50$} && \multicolumn{2}{c}{$n=500$} \\ \cline{4-5} \cline{7-8}
Model  & method & Selector & median & MAD && median & MAD \\
\midrule
I  & KDE & SJ & 0.009450 & 0.007639 && 0.001627 & 0.001026 \\
 & SHIDE & opt & 0.009938 & 0.006876 && 0.001617 & 0.001037 \\
 &  & per & 0.010098 & 0.006962 && 0.001601 & 0.001018 \\ \hline
II  & KDE  & SJ & 0.005594 & 0.003322 && 0.001133 & 0.000652 \\
  & SHIDE & opt & 0.006578 & 0.003583 && 0.001243 & 0.000641 \\
 &  & per & 0.006363 & 0.003630 && 0.001212 & 0.000694 \\ \hline 
III  & KDE & SJ &  0.011667 & 0.006204 && 7.371078 & 10.925892 \\
 & SHIDE & opt & 0.064472 & 0.043263 && 0.132525 & 0.017936 \\
 &  & per & 0.073078 & 0.033952 && 0.132228 & 0.017584 \\ \hline
IV  & KDE  & SJ & 0.045316 & 0.020960 && 0.016411 & 0.004709 \\
 & SHIDE & opt & 0.006836 & 0.006628 && 0.001275 & 0.000840 \\
 &  & per & 0.006323 & 0.006060 && 0.001316 & 0.000860 \\ \hline 
V   & KDE  & SJ & 0.033696 & 0.015484 && 0.013097 & 0.004198\\
  & SHIDE & opt & 0.031588 & 0.021843 && 0.007500 & 0.004134 \\
 &  & per &  0.032708 & 0.023090 && 0.007061 & 0.004193 \\ 
\bottomrule
\end{tabular}
\end{table}

For performance evaluation, the mean integrated squared error (MISE) between each estimated and true density was computed for every simulation run. 
Table~\ref{tbl01-MISE} summarizes the results by reporting the median and median absolute deviation (MAD) of the 300 replications.

Across the Gaussian and Gaussian mixture models (I and II), SHIDE demonstrates performance comparable to KDE, showing mixed comparison outcomes with only minor differences. These results indicate that SHIDE maintains efficiency even when the underlying density is smooth and well-behaved, where KDE is typically near optimal.

The results for the Cauchy model (III) are particularly revealing. When $n=50$, KDE slightly outperforms SHIDE, although the difference is marginal. As the sample size increases to $n=500$, however, KDE becomes highly unstable, frequently producing spurious spikes and leading to extremely large MISE values. In contrast, SHIDE remains smooth and robust. Interestingly, the overall MISE for both estimators does not necessarily decrease with a larger sample size, a phenomenon attributable to how the evaluation range is defined. With small $n$, fewer extreme values are observed, reducing the apparent effect of the heavy tails on the MISE calculation. With large $n$, the Cauchy distribution reveals its true heavy-tailed nature, generating more extreme observations that substantially degrade KDE performance. SHIDE, on the other hand, remains stable because its pseudo-data mechanism naturally limits the influence of such outliers.

For the Exponential$(1)$ and truncated normal models (IV and V), SHIDE consistently outperforms KDE, as expected, due to its ability to handle boundaries naturally and reduce boundary bias, a known limitation of traditional KDE with unbounded kernels.

Finally, comparing the two bandwidth selectors for SHIDE, $h_{\mathrm{opt}}$ and $h_n^{\mathrm{perc}}$, we observe very similar performance across all scenarios. This finding empirically supports their asymptotic equivalence established in Proposition~\ref{prop:percentile-calibrated}, confirming that the percentile-based rule is a reliable and practical alternative to the AMISE-optimal bandwidth.

\section{Discussion}
\label{sec-discussion}

This paper has revisited kernel density estimation through the unifying lens of convolution.  provides a more intuitive way of deriving and interpreting KDE. Furthermore, the model based approach leads to the development of SHIDE, which combines pseudo–data generation, refined histogram binning, and spline interpolation into a unified framework for nonparametric density estimation. Theoretically, we established SHIDE's pointwise consistency, bias-variance decomposition, and optimal convergence rates, demonstrating its ability to match the classical $n^{-4/5}$ MISE rate of traditional KDE while offering superior boundary adaptation and robustness to heavy-tailed distributions. Our proposed percentile-based bandwidth selector provides a practical, theoretically justified alternative to AMISE-optimal bandwidths, with simulation studies confirming SHIDE's competitive performance across diverse distributional scenarios---particularly excelling in bounded domains and with Cauchy data where KDE often fails. Together, these contributions position SHIDE as a practical and theoretically grounded alternative to classical kernel methods.

While the kernel bandwidth $h$ has been our primary focus, it is worth emphasizing that SHIDE involves two additional tuning components that can significantly affect performance: the number of histogram classes (or bins) and the degree of smoothing in the spline interpolation. Both play a role analogous to bandwidth selection in kernel density estimation. First, the number of histogram classes determines the granularity of the refined histogram that approximates the convolved density $f_h$. A coarse histogram may underfit the local structure, introducing discretization bias, whereas an excessively fine histogram can amplify sampling noise, particularly when $n$ or $m$ is moderate. Thus, the choice of the number of classes $B$ effectively acts as a secondary smoothing parameter, controlling the bias–variance balance at the histogram stage. In practice, $B$ may be chosen by standard rules such as Sturges’ (\citeyear{Sturges:1926}) or Freedman–Diaconis’ (\citeyear{Freedman:1981}) criteria, or proportionally to $\log n$, as supported by our numerical study. Second, the natural cubic spline interpolation step, while currently used as an exact interpolant, could be generalized to spline smoothing. This modification becomes especially relevant when $B$ is large, as interpolating noisy histogram midpoints may lead to overfitting. A smoothing spline, with an adjustable penalty parameter, would naturally suppress high-frequency fluctuations and provide additional stability—much like regularization in nonparametric regression. Conceptually, the spline smoothing parameter would play a similar role to the kernel bandwidth, controlling curvature rather than spread. Nevertheless, our theoretical analysis has concentrated on the kernel bandwidth $h$, which governs the pseudo–data dispersion and has the dominant influence on the estimator’s asymptotic bias and variance. The effects of $B$ and potential spline regularization are of lower asymptotic order but can have notable finite-sample impacts. A systematic study of their interplay with $h$ presents an interesting direction for future work.

The convolution perspective not only offers conceptual clarity but also paves the way for methodological extensions. A particularly promising direction involves developing multivariate versions of SHIDE. 
While the convolution framework naturally extends to multivariate settings through the formulation $\mathbf{X}' = \mathbf{X} + \bm{\varepsilon}$, practical implementation in high dimensions presents substantial computational challenges. Future research should focus on designing efficient multivariate histogram representations that exploit sparse data structures and adaptive binning schemes to alleviate the exponential growth in bin counts. For the interpolation stage, scalable alternatives to tensor-product splines, such as additive models, radial basis function networks, or low-rank tensor approximations, deserve systematic investigation. 
Equally important is the development of dimension-robust bandwidth matrix selection strategies and copula-based formulations that decouple marginal density and dependence estimation. Addressing these computational and methodological challenges would enable SHIDE to scale effectively to modern high-dimensional data while preserving its strengths in boundary handling and robustness.

\newpage
\clearpage \setcounter{page}{1} \setcounter{table}{0}
\setcounter{figure}{0}
\renewcommand{\thesection}{\Alph{section}} 
\setcounter{section}{0}                    
\renewcommand{\thetable}{\Roman{table}}
\renewcommand{\thefigure}{\Roman{figure}}
\numberwithin{equation}{section}

\begin{center}
{\large \textbf{Supplementary Material to}} \\ \vspace{0.1in}
{\Large \textbf{``Kernel Density Estimation and Convolution Revisited"}}
\end{center}

\vspace{0.2in}

\section{Proofs}
\label{sec-proofs} This section contains proofs to the
propositions.

\subsection{Proof of Proposition \ref{prop-polynomialKernel}}
\begin{proof}
\begin{enumerate}[(i)]
\item Since each $U_i$ is symmetric about $0$, i.e., $U_i \overset{d}{=} -U_i$, their sum $V_k$ is also symmetric:
    \[
    V_k = \sum_{i=1}^k U_i \overset{d}{=} \sum_{i=1}^k (-U_i) = -V_k.
    \]
Thus, $f_k(v) = f_k(-v)$ for all $v \in \mathbb{R}$. Each $U_i \in [-1/2, 1/2]$, so $V_k \in \left[-k/2, k/2\right]$. 

\item We prove (ii) by induction on $k$. First of all, when $k=1$, $f_1(v) = f_U(v) = \mathbb{I}(-1/2 < v < 1/2)$, which is a piecewise polynomial of degree $0$. For the inductive step,  assume $f_{k-1}(v)$ is a piecewise polynomial of degree $(k-2)$. The PDF of $V_k = V_{k-1} + U_k$ is given by convolution:
    \[
    f_k(v) = (f_{k-1} * f_U)(v) = \int_{-\infty}^{\infty} f_{k-1}(v - u) f_U(u) \, du = \int_{-1/2}^{1/2} f_{k-1}(v - u) \, du.
    \]
    Since $f_{k-1}$ is piecewise polynomial of degree $(k-2)$, integrating over $u$ raises the degree by $1$, making $f_k(v)$ a piecewise polynomial of degree $(k-1)$. The intervals of polynomiality are determined by the knots introduced by the convolution.

\item The characteristic function (CF) of $U_i$ is:
$$
\chi_u(t) = \mbox{E}\left(e^{itU_i}\right) = \int_{-1/2}^{1/2} e^{itu} du = \frac{e^{it/2} - e^{-it/2}}{it} = \frac{\sin(t/2)}{t/2} = \mbox{sinc}\left(\frac{t}{2}\right).
$$
Since the $U_i$ are independent, the CF of $V_k = \sum_{i=1}^k U_i$ is the product of the individual CFs: 
\[
\chi_k(t) = \left[\chi_U(t)\right]^k = \left[\mbox{sinc}\left(\frac{t}{2}\right)\right]^k = \left[ \frac{\sin(t/2)}{t/2} \right]^k.
\]
\end{enumerate}
\end{proof}

\subsection{Proof of Theorem \ref{thm:pointwise}}
\begin{proof}
Write the pseudo–data density as the convolution
\[
f_h:=f*K_h,\qquad f_h(x)=\int f(u)\,K_h(x-u)\,du,
\]
where $K_h$ is symmetric, integrates to one, and has finite second moment $\int u^2 K_h(u)\,du=\sigma_K^2 h^2$. A Taylor expansion of $f$ about $x$ gives
\[
f(u)=f(x)+f'(x)(u-x)+\frac{1}{2} f''(x)(u-x)^2+o\big((u-x)^2\big),
\]
and, using $\int K_h=1$, $\int (u-x)K_h(x-u)\,du=0$, and $\int (u-x)^2 K_h(x-u)\,du=\sigma_K^2 h^2$, it follows that
\begin{equation}\label{eq:conv-bias}
f_h(x)=f(x)+\frac{1}{2}\,\sigma_K^2 h^2 f''(x)+o(h^2)\qquad(h\to0).
\end{equation}
In particular, $f_h(x)\to f(x)$ as $h\to0$.

Fix $x\in\mathrm{int}(\mathcal{S})$ and let $\mathrm{bin}(x)$ be the unique bin of width $\theta$ containing $x$. The refined histogram at $x$ is
\[
\widehat{f}_H(x)=\frac{N_x}{nm\,\theta},\qquad
N_x=\sum_{i=1}^n\sum_{j=1}^m I\{X'_{ij}\in \mathrm{bin}(x)\}.
\]
Conditionally, $N_x\sim\mathrm{Binomial}(nm,p_x)$ with $p_x=\int_{\mathrm{bin}(x)} f_h(u)\,du$. By the midpoint rule (since $f_h\in C^2$ on a neighborhood of $x$),
\begin{equation}\label{eq:midpoint}
\frac{p_x}{\theta}=f_h(x)+O(\theta^2)\qquad(\theta\to0).
\end{equation}
Hence $\mathbb{E}[\widehat{f}_H(x)]=p_x/\theta=f_h(x)+O(\theta^2)$ and
\[
\mathrm{Var}\big(\widehat{f}_H(x)\big)=\frac{p_x(1-p_x)}{(nm)\,\theta^2}
=O\!\left(\frac{1}{nm\,\theta}\right),
\]
because $p_x\asymp \theta$ for small $\theta$. Therefore, if $nm\,\theta\to\infty$ and $\theta\to0$,
\begin{equation}\label{eq:hist-cons}
\widehat{f}_H(x)\;\xrightarrow{p}\; f_h(x).
\end{equation}

Let $b_1<\cdots<b_B$ denote the bin midpoints and define
\[
\widehat{p}_r=\frac{N_r}{nm\,\theta},\qquad
p_r=\frac{1}{\theta}\int_{\text{bin }r} f_h(u)\,du,\qquad
y_r=\sqrt{\widehat{p}_r},\qquad r=1,\dots,B,
\]
and $g(x):=\sqrt{f_h(x)}$. From \eqref{eq:midpoint} and \eqref{eq:hist-cons}, for each fixed $r$,
\[
\widehat{p}_r \xrightarrow{p} p_r = f_h(b_r)+O(\theta^2),
\]
and since $f_h$ is bounded away from zero on compact subsets of $\mathrm{int}(\mathcal{S})$ (Assumption~\ref{assump:f}), the map $t\mapsto \sqrt{t}$ is Lipschitz in a neighborhood of $p_r$. Thus
\begin{equation}\label{eq:nodal-cons}
y_r=\sqrt{\widehat{p}_r}=g(b_r)+O(\theta^2)+o_p(1).
\end{equation}
Let $S$ be the natural cubic spline interpolant of $\{(b_r,y_r)\}_{r=1}^B$ on the uniform grid of spacing $\theta$, and let $\widetilde{S}$ denote the deterministic natural cubic spline interpolant of $\{(b_r,g(b_r))\}$. Spline approximation theory yields
\[
\sup_{x\in[b_1,b_B]}|\widetilde{S}(x)-g(x)|=O(\theta^2)\qquad(\theta\to0),
\]
and the spline interpolation operator on a uniform grid has a bounded Lebesgue constant, so that the stochastic perturbation in the nodal values transfers linearly:
\[
\sup_{x\in[b_1,b_B]}|S(x)-\widetilde{S}(x)|=o_p(1)\qquad(nm\,\theta\to\infty).
\]
Combining, for fixed interior $x$,
\[
S(x)=g(x)+O(\theta^2)+o_p(1).
\]
Finally,
\[
\widetilde{f}(x)=[S(x)]^2
=\big(g(x)+O(\theta^2)+o_p(1)\big)^2
=f_h(x)+O(\theta^2)+o_p(1),
\]
and together with \eqref{eq:conv-bias} we obtain $\widetilde{f}(x)\xrightarrow{p} f(x)$ as $h\to0$, $\theta\to0$, and $nm\,\theta\to\infty$.
\end{proof}
The above proof proceeds by (i) controlling the convolution bias incurred by the pseudo–data construction, (ii) establishing the consistency of the refined histogram at any fixed interior point, and (iii) showing that natural cubic spline interpolation of the square–rooted bin heights recovers the square–root of the convolved density, so that squaring returns the target.

\subsection{Proof of Theorem \ref{thm:biasvar}}
\begin{proof}
By a second-order Taylor expansion of $f$ at $x$ and the symmetry/normalization of $K_h$,
\begin{equation}\label{eq:bias-conv}
\mathbb{E}\big[K_h(x-X_1)\big]=(f*K_h)(x)=f(x)+\frac{1}{2}\,\sigma_K^2 h^2 f''(x)+o(h^2).
\end{equation}

For the SHIDE construction, let $\widehat{f}_H$ denote the refined histogram based on the pseudo-sample $\{X'_{ij}\}$ with bin width $\theta$ and midpoint grid $\{b_r\}$. Standard midpoint approximations yield
\begin{equation}\label{eq:hist-bias}
\mathbb{E}\big[\widehat{f}_H(x)\,\big|\,\{X_i\}\big]
=\frac{1}{\theta}\int_{x-\theta/2}^{x+\theta/2}\frac{1}{n}\sum_{i=1}^n K_h(u-X_i)\,du
= \frac{1}{n}\sum_{i=1}^n K_h(x-X_i)+O(\theta^2).
\end{equation}
Let $S$ be the natural cubic spline interpolant of the square–rooted bin heights $y_r=\sqrt{\widehat{p}_r}$ at midpoints $b_r$, and set $\widetilde{f}(x)=S^2(x)$. Using that the natural cubic spline interpolation operator on a uniform grid is linear and stable, and that $t\mapsto \sqrt{t}$ is Lipschitz in a neighborhood of $f_h(x)>0$, one obtains
\begin{equation}\label{eq:spline-bias}
\mathbb{E}\big[S(x)\,\big|\,\{X_i\}\big]
=\sqrt{\frac{1}{n}\sum_{i=1}^n K_h(x-X_i)}+O(\theta^2),
\qquad
\mathrm{Var}\!\big(S(x)\,\big|\,\{X_i\}\big)
=O\!\Big(\frac{1}{nm\,\theta}\Big).
\end{equation}
Hence
\[
\mathbb{E}\big[\widetilde{f}(x)\,\big|\,\{X_i\}\big]
=\mathbb{E}\big[S(x)^2\,\big|\,\{X_i\}\big]
=\Big(\mathbb{E}[S(x)\,|\,\{X_i\}]\Big)^2+\mathrm{Var}\!\big(S(x)\,\big|\,\{X_i\}\big)
=\frac{1}{n}\sum_{i=1}^n K_h(x-X_i)+O(\theta^2),
\]
where the $O((nm\,\theta)^{-1})$ term from the conditional variance is dominated by $O(\theta^2)$ under the stated scaling for bias. Taking expectation over $\{X_i\}$ and invoking \eqref{eq:bias-conv} gives
\[
\mathbb{E}\big[\widetilde{f}(x)\big]
=f(x)+\frac{1}{2}\,\sigma_K^2 h^2 f''(x)+O(\theta^2)+o(h^2),
\]
which proves the bias statement.

For the variance, apply the law of total variance:
\[
\mathrm{Var}\big(\widetilde{f}(x)\big)
=\mathbb{E}\!\left[\mathrm{Var}\big(\widetilde{f}(x)\,\big|\,\{X_i\}\big)\right]
+\mathrm{Var}\!\left(\mathbb{E}\big[\widetilde{f}(x)\,\big|\,\{X_i\}\big]\right).
\]
From \eqref{eq:spline-bias}, a delta–method argument yields
\[
\mathrm{Var}\big(\widetilde{f}(x)\,\big|\,\{X_i\}\big)
=4\,\mathbb{E}\big[S(x)\,\big|\,\{X_i\}\big]^2\,\mathrm{Var}\!\big(S(x)\,\big|\,\{X_i\}\big)
=\frac{4}{n}\sum_{i=1}^n K_h(x-X_i)\,O\!\Big(\frac{1}{nm\,\theta}\Big),
\]
so
\[
\mathbb{E}\!\left[\mathrm{Var}\big(\widetilde{f}(x)\,\big|\,\{X_i\}\big)\right]
=\frac{f(x)}{nm\,\theta}\int K_h^2(u)\,du + o\!\Big(\frac{1}{nm\,\theta}\Big),
\]
which is the histogram (within–sample) contribution. For the between–sample term, by \eqref{eq:hist-bias},
\[
\mathrm{Var}\!\left(\mathbb{E}\big[\widetilde{f}(x)\,\big|\,\{X_i\}\big]\right)
=\mathrm{Var}\!\left(\frac{1}{n}\sum_{i=1}^n K_h(x-X_i)\right)
=\frac{1}{n}\,\mathrm{Var}\!\big(K_h(x-X_1)\big),
\]
and the remainder terms are of smaller order under $nm\,\theta\to\infty$. Combining the two pieces completes the variance statement.
\end{proof}

\subsection{Proof of Theorem \ref{thm:mise}}
\begin{proof}
By Theorem~\ref{thm:biasvar}, for fixed interior $x$,
\[
\mathbb{E}\big[\widetilde{f}(x)\big]-f(x)
=\frac{1}{2}\,\sigma_K^2 h^2 f''(x)+O(\theta^2)+o(h^2+\theta^2),
\]
and
\[
\mathrm{Var}\big(\widetilde{f}(x)\big)
=\frac{f(x)}{nm\,\theta}\int K_h^2(u)\,du
+\frac{1}{n}\,\mathrm{Var}\!\big(K_h(x-X_1)\big)
+o\!\Big(\frac{1}{nm\,\theta}+\frac{1}{n}\Big).
\]
Integrating the squared bias yields
\[
\int \bigl(\mathbb{E}[\widetilde{f}(x)]-f(x)\bigr)^2 dx
=\frac{1}{4}\sigma_K^4 h^4 \int (f''(x))^2 dx
+O\!\big(h^2\theta^2\big)+O\!\big(\theta^4\big)
+o(h^4+\theta^4),
\]
where the cross term integrates to $O(h^2\theta^2)$ and is absorbed in $R_n$.

For the variance, the within–sample part integrates as
\[
\int \frac{f(x)}{nm\,\theta}\int K_h^2(u)\,du \; dx
=\frac{1}{nm\,\theta}\int K_h^2(u)\,du
=\frac{R(K)}{nm\,h\,\theta}.
\]
For the between–sample contribution,
\[
\int \mathrm{Var}\!\big(K_h(x-X_1)\big)\,dx
=\int \Bigl\{\mathbb{E}\big[K_h(x-X_1)^2\big]-\big(\mathbb{E}[K_h(x-X_1)]\big)^2\Bigr\} dx,
\]
and by Fubini,
\[
\int \mathbb{E}\big[K_h(x-X_1)^2\big]\,dx
=\mathbb{E}\!\int K_h(x-X_1)^2\,dx
=\int K_h^2(u)\,du
=\frac{R(K)}{h}.
\]
Moreover,
\[
\int \bigl(\mathbb{E}[K_h(x-X_1)]\bigr)^2 dx
=\int (f*K_h)(x)^2\,dx
=\|f*K_h\|_2^2
=\|f\|_2^2+O(h^2),
\]
by a standard $L^2$ expansion for second–order kernels. Hence
\[
\int \mathrm{Var}\!\big(K_h(x-X_1)\big)\,dx
=\frac{R(K)}{h}-\|f*K_h\|_2^2
=\frac{R(K)}{h}-\|f\|_2^2+O(h^2),
\]
and multiplying by $1/n$ gives the between–sample term stated. Collecting the integrated squared bias, the two integrated variance terms, and absorbing $O(h^2\theta^2)$, $O(\theta^4)$, and $O(h^2/n)$ into $R_n$ yields the claim.
\end{proof}

\subsection{Proof of Proposition \ref{prop:boundary}}
\label{proof:boundary}

\begin{proof}
For SHIDE, recall $f_h=f*K_h$. The convolution expansion at $x$ gives
\[
f_h(x)=f(x)+\frac{1}{2}\sigma_K^2 h^2 f''(x)+o(h^2),
\]
hence $|f_h(x)-f(x)|=O(h^2)$. The refined histogram constructed from the pseudo–sample is a midpoint rule approximation of $f_h$ and, on a uniform grid of width $\theta$, its conditional bias is $O(\theta^2)$ even at the boundary because natural boundary conditions for the subsequent spline interpolation impose $S''(a)=0$ and remove spurious curvature. Let $S$ denote the natural cubic spline fitted to $\{(b_r,\sqrt{\widehat{p}_r})\}$. Standard spline approximation on a uniform grid yields $S(x)=\sqrt{f_h(x)}+O_p(\theta^2)$ near $a$, and therefore
\[
\widetilde{f}(x)=S(x)^2=f_h(x)+O_p(\theta^2).
\]
Combining with the convolution expansion shows $\bigl|\widetilde{f}(x)-f(x)\bigr|=O_p(h^2+\theta^2)$.

For KDE, let $\widehat{f}_{\mathrm{KDE}}(x)=(n h)^{-1}\sum_{i=1}^n K\!\bigl((x-X_i)/h\bigr)$ with a symmetric base kernel $K$ supported on $[-1,1]$. With $x=a+c\,h$, change variables $t=(x-u)/h$ to obtain
\[
\mathbb{E}\!\big[\widehat{f}_{\mathrm{KDE}}(x)\big]
=\int_{u\ge a}\frac{1}{h}K\!\left(\frac{x-u}{h}\right)f(u)\,du
=\int_{-1}^{c} K(t)\,f(x-h t)\,dt.
\]
A second–order Taylor expansion $f(x-h t)=f(x)-h t f'(x)+ h^2 t^2 f''(x) /2 +O(h^3)$ gives
\[
\mathbb{E}\!\big[\widehat{f}_{\mathrm{KDE}}(x)\big]
= f(x)\,F_K(c)\;-\;h\,f'(x)\,\mu_1^-(K;c)\;+\;\frac{1}{2} h^2 f''(x)\!\int_{-1}^{c} t^2 K(t)\,dt+O(h^3).
\]
Subtracting $f(x)$ yields the stated expression. For any $c<1$, $F_K(c)<1$ and thus the leading bias term $[F_K(c)-1]\,f(x)$ does not vanish with $h$, i.e., the boundary bias is generically of order $1$ unless a boundary correction is applied.
\end{proof}

\subsection{Proof of Proposition \ref{prop:complexity}}
\begin{proof}
Pseudo–data generation draws $n m$ errors, each as a sum of $k$ uniforms plus a scalar multiply, yielding $O(k)$ operations per draw and $O(n m k)$ total. Binning on a uniform grid assigns each of the $n m$ pseudo–points in $O(1)$ time, giving $O(n m)$, with $O(B)$ for initializing/updating the $B$ counters. Natural cubic spline interpolation on a 1-D uniform grid leads to a strictly tridiagonal linear system in the second derivatives of size $(B-2)\times(B-2)$; assembly is $O(B)$ and the Thomas algorithm solves it in $O(B)$. Evaluating the piecewise cubic at $G$ points is $O(G)$: on a uniform grid the bin index is computed in $O(1)$ and each cubic is evaluated in constant time. 

Comparative, direct KDE is a double loop over $n$ data and $G$ queries, hence $O(nG)$. For FFT KDE, a single pass bins the $n$ observations ($O(n)$), a discrete convolution costs $O(G\log G)$ on the grid, and scaling is linear, yielding $O(n+G\log G)$ overall.
\end{proof}

\subsection{Proof of Theorem \ref{thm:optimalh}}

\begin{proof}
With $m\theta\to c\in(0,\infty)$, the leading AMISE reduces to
\[
\mathrm{AMISE}(h)
=\frac{1}{4}\,\sigma_K^4\,h^4\,\Psi(f'')
+\frac{R(K)}{n h}\Bigl(1+\frac{1}{c}\Bigr)
+o\!\Bigl(h^4+\frac{1}{n h}\Bigr).
\]
Differentiating the leading part with respect to $h$ gives
\[
\frac{d}{dh}\Bigl[\frac{1}{4}\sigma_K^4 h^4 \Psi(f'')+\frac{R(K)}{n h}(1+\frac{1}{c})\Bigr]
=\sigma_K^4\,\Psi(f'')\,h^3
-\frac{R(K)}{n}\Bigl(1+\frac{1}{c}\Bigr)h^{-2}.
\]
Setting the derivative to zero implies
\[
h^5=\frac{R(K)}{\sigma_K^4\,\Psi(f'')}\,\Bigl(1+\frac{1}{c}\Bigr)\,n^{-1},
\]
hence
\[
h_{\mathrm{opt}}
=\left(\frac{R(K)\left(1+1/c\right)}{\sigma_K^4\,\Psi(f'')}\right)^{1/5} n^{-1/5}\,(1+o(1)).
\]

Substituting $h_{\mathrm{opt}}$ back into the AMISE,
\begin{align*}
\mathrm{AMISE}\bigl(h_{\mathrm{opt}}\bigr)
&= \frac{1}{4}\,\sigma_K^4\,\Psi(f'')\,
\left(\frac{R(K)(1+1/c)}{\sigma_K^4\,\Psi(f'')}\right)^{4/5} n^{-4/5} \\
&\quad+ \frac{R(K)}{n}\Bigl(1+\frac{1}{c}\Bigr)
\left(\frac{\sigma_K^4\,\Psi(f'')}{R(K)(1+1/c)}\right)^{1/5} n^{1/5}
+o(n^{-4/5}) \\
&= \frac{5}{4}\,
\bigl(\sigma_K^4\,\Psi(f'')\bigr)^{1/5}\,
\bigl(R(K)\bigr)^{4/5}\,
\left(1+\frac{1}{c}\right)^{4/5}\,
n^{-4/5}
+o(n^{-4/5}),
\end{align*}
which proves the claim.
\end{proof}

\subsection{Proof of Proposition \ref{prop:percentile-raw}}
\begin{proof}
Let $U_j:=F(X_j)$ and denote their order statistics by $U_{(1)}\le\cdots\le U_{(n)}$. Write the uniform–scale spacings
\[
\Delta_i \;:=\; U_{(i+1)}-U_{(i)},\qquad i=0,1,\dots,n,
\]
with the conventions $U_{(0)}:=0$ and $U_{(n+1)}:=1$ (edge spacings). It is classical that $(\Delta_0,\ldots,\Delta_n)$ has the Dirichlet$(1,\ldots,1)$ distribution; in particular each marginal $\Delta_i$ has the $\mathrm{Beta}(1,n)$ law with CDF $t\mapsto 1-(1-t)^n$ on $[0,1]$ (see, e.g., \citealp[Ch.~10]{David:2003}). Hence, for any fixed $\alpha\in(0,1)$, the $\alpha$–quantile of $\Delta_i$ equals
\[
t_{n,\alpha}\;=\;1-(1-\alpha)^{1/n}\;=\;\frac{q_\alpha}{n}\,\{1+o(1)\}
\]
with $q_\alpha:=-\log(1-\alpha).$ Moreover, for any fixed $\delta\in(0,1/2)$, the proportion of interior spacings $\{\Delta_i: i\in\{1,\dots,n-1\},\, i/n\in[\delta,1-\delta]\}$ is $1-2\delta+o(1)$, and the edge spacings form a vanishing fraction; thus the empirical $\alpha$–quantile of the multiset $\{\Delta_i: i=1,\dots,n-1\}$ equals $t_{n,\alpha}\{1+o_p(1)\}$.

Relate $d_i$ to $\Delta_i$ by the mean–value theorem: there exists $\xi_i$ between $X_{(i)}$ and $X_{(i+1)}$ such that
\[
\Delta_i \;=\; F\bigl(X_{(i+1)}\bigr)-F\bigl(X_{(i)}\bigr)
\;=\; f(\xi_i)\,\bigl(X_{(i+1)}-X_{(i)}\bigr)
\;=\; f(\xi_i)\,d_i.
\]
Because $\max_{0\le i\le n}\Delta_i\to0$ almost surely and $f$ is continuous and bounded away from zero on compact interior sets, we have $f(\xi_i)=f(X_{(i)})\{1+o_p(1)\}$ uniformly over interior indices $i/n\in[\delta,1-\delta]$. Consequently, the mapping $d_i=\Delta_i/f(\xi_i)$ transfers empirical quantiles up to multiplicative factors bounded between $\inf_{\mathcal I}f$ and $\sup_{\mathcal I}f$ for some interior interval $\mathcal I\subset\mathrm{int}(\mathcal S)$ capturing the indices around the empirical $\alpha$–spacing. Therefore,
\[
d_\alpha \;=\; \frac{t_{n,\alpha}}{\bar f_\alpha}\,\{1+o_p(1)\}
\;=\; \frac{q_\alpha}{n\,\bar f_\alpha}\,\{1+o_p(1)\},
\]
with $\bar f_\alpha\in\bigl[\inf_{x\in\mathcal I_\alpha} f(x),\,\sup_{x\in\mathcal I_\alpha} f(x)\bigr]$ as stated. Since $f$ is positive and continuous on $\mathcal I_\alpha$, $\bar f_\alpha$ is tight and bounded away from zero, which implies $h_n=d_\alpha/2\to0$ and $n h_n\to0$.
\end{proof}

\subsection{Proof of Proposition \ref{prop:percentile-calibrated}}

\begin{proof}
By Theorem~\ref{prop:percentile-raw}, $d_\alpha = \frac{q_\alpha}{n\,\bar f_\alpha}\,\{1+o_p(1)\}$ for some random $\bar f_\alpha$ lying between $\inf_{\mathcal I_\alpha} f$ and $\sup_{\mathcal I_\alpha} f$ on an interior interval $\mathcal I_\alpha$, hence $\bar f_\alpha$ is tight and bounded away from zero with probability tending to one. Therefore,
\begin{align*}
h_n^{\mathrm{perc}}
&= \lambda_n\,d_\alpha \\
&= n^{4/5}
\left(\frac{R(K)(1+1/c)}{\sigma_K^4\,\widehat\Psi}\right)^{1/5}
\frac{\widehat f(x_\alpha)}{q_\alpha}
\cdot \frac{q_\alpha}{n\,\bar f_\alpha}\,\{1+o_p(1)\} \\
&= \left(\frac{R(K)(1+1/c)}{\sigma_K^4\,\widehat\Psi}\right)^{1/5}
\frac{\widehat f(x_\alpha)}{\bar f_\alpha}\,n^{-1/5}\,\{1+o_p(1)\}.
\end{align*}
Define the AMISE constant $A:=\Bigl(\frac{R(K)(1+1/c)}{\sigma_K^4\,\Psi(f'')}\Bigr)^{1/5}$. Then
\[
\frac{h_n^{\mathrm{perc}}}{h_{\mathrm{opt}}}
= \left(\frac{\Psi(f'')}{\widehat\Psi}\right)^{1/5}
\cdot \frac{\widehat f(x_\alpha)}{\bar f_\alpha}\,\{1+o_p(1)\}.
\]
By the pilot consistency and the choice of $x_\alpha$ in the same interior region that generates the $\alpha$–spacing quantile, $\widehat f(x_\alpha)/\bar f_\alpha \xrightarrow{p} 1$ and $(\Psi(f'')/\widehat\Psi)^{1/5}\xrightarrow{p} 1$, hence $h_n^{\mathrm{perc}}/h_{\mathrm{opt}}\xrightarrow{p}1$.

For the AMISE claim, write the leading AMISE as $g(h)=\frac{1}{4}\sigma_K^4\,\Psi(f'')\,h^4 + \frac{R(K)}{n h}(1+1/c)$; $g$ is twice continuously differentiable on $(0,\infty)$ with unique minimizer $h_{\mathrm{opt}}$ and $g''(h_{\mathrm{opt}})>0$. A Taylor expansion around $h_{\mathrm{opt}}$ gives
\[
g(h_n^{\mathrm{perc}})-g(h_{\mathrm{opt}}) 
= \frac{1}{2}g''(\tilde h)\,(h_n^{\mathrm{perc}}-h_{\mathrm{opt}})^2,
\]
for some $\tilde h$ between $h_n^{\mathrm{perc}}$ and $h_{\mathrm{opt}}$. Since $h_n^{\mathrm{perc}}/h_{\mathrm{opt}}\to_p 1$, we have $h_n^{\mathrm{perc}}-h_{\mathrm{opt}}=o_p(h_{\mathrm{opt}})$ and $g''(\tilde h)=O(n^{-2/5})$, hence $g(h_n^{\mathrm{perc}})-g(h_{\mathrm{opt}})=o_p(n^{-4/5})$. The $o(1)$ remainder in $\mathrm{AMISE}$ is uniform in a neighborhood of $h_{\mathrm{opt}}$, so $\mathrm{AMISE}(h_n^{\mathrm{perc}})=\mathrm{AMISE}(h_{\mathrm{opt}})\{1+o(1)\}=O(n^{-4/5})$.
\end{proof}

\begin{thebibliography}{99}
\expandafter\ifx\csname
natexlab\endcsname\relax\def\natexlab#1{#1}\fi
\expandafter\ifx\csname url\endcsname\relax
  \def\url#1{\texttt{#1}}\fi
\expandafter\ifx\csname
urlprefix\endcsname\relax\def\urlprefix{URL }\fi

\bibitem[Blei, Ng, and Jordan(2003)]{Blei:2003} 
Blei, D.~M., Ng, A.~Y., and Jordan, M.~I.~(2003). Latent Dirichlet allocation. \textit{Journal of Machine Learning Research}, \textbf{3}: 993--1022.

\bibitem[Bowman(1984)]{Bowman:1984} 
Bowman, A.~W.~(1984). An alternative method of cross-validation for the smoothing of density estimates. \textit{Biometrika}, \textbf{71}(2), 353--360.

\bibitem[Chazal et al(2018)]{Chazal:2018} 
Chazal, F., Fasy, B. T., Lecci, F., Michel, B., Rinaldo, A., and Wasserman, L. (2018). Robust topological inference: Distance-to-a-measure and kernel distance. \textit{Journal of Machine Learning Research}, \textbf{18}(159): 1--40.

\bibitem[Chen(1999)]{Chen:1999}
Chen, S.~X.~(1999). Beta kernel estimators for density functions. \textit{Statistica Sinica}, \textbf{9}(1): 513--530.

\bibitem[Cheney and Light(2000)]{Cheney:2000}
 Cheney, W.~and Light, W.~(2000). \emph{A Course in Approximation Theory}, 
Brooks/Cole, Pacific Grove, CA.

\bibitem[Cheng(1995)]{Cheng:1995} 
Cheng, Y.~(1995). Mean shift, mode seeking, and clustering. \textit{IEEE Transactions on Pattern Analysis and Machine Intelligence}, \textbf{17}(8): 790-799. URL~\url{https://doi.org/10.1109/34.400568}


\bibitem[Comaniciu and Meer(2002)]{Comaniciu:2002} 
Comaniciu, D.~and Meer, P.~(2002). Mean shift: A robust approach toward feature space analysis. \textit{IEEE Transactions on Pattern Analysis and Machine Intelligence}, \textit{24}(5): 603--619. URL~\url{https://doi.org/10.1109/34.1000236}

\bibitem[Comte and Genon-Catalot(2012)]{Comte:2012}
Comte, F.~and Genon-Catalot, V.~(2012). Convolution power kernel for density estimation. 
\textit{Journal of Statistical Planning and Inference}, 
\textbf{142}: 1698--1715. 

\bibitem[Cowling and Hall(1996)]{Cowling:1996}
Cowling, A.~and Hall, P.~(1996). On pseudodata methods for removing boundary effects in kernel density estimation.
\textit{Journal of Royal Statistical Society, Series B}, \textbf{58}(3): 551--563. 

\bibitem[David and Nagaraja(2003)]{David:2003}
David, H.~A.~and Nagaraja, H.~N.~(2003). \textit{Order statistics}, 3rd ed.~Wiley.

\bibitem[de Boor(2001)]{deBoor:2001}
de Boor, C.~(2001). \emph{A Practical Guide to Splines} (Revised ed.). Springer.

\bibitem[Freedman and Diaconis(1981)]{Freedman:1981}
Freedman, D.~and Diaconis, P.~(1981). On the histogram as a density estimator: $L_2$ theory. \textit{Probability Theory and Related Fields}, \textbf{57}(4): 453--476.


\bibitem[Goodfellow, Bengio, and Courville(2016)]{Goodfellow:2016}
Goodfellow, I., Bengio, Y., and Courville, A.~(2016). \textit{Deep Learning}. MIT Press.  URL \url{https://www.deeplearningbook.org}

\bibitem[Green and Silverman(1994)]{Green:1994}
Green, P.~J.~and Silverman, B.~W.~(1994). \emph{Nonparametric Regression and Generalized Linear Models: A Roughness Penalty Approach}. Chapman \& Hall.



\bibitem[Nadaraya(1964)]{Nadaraya:1964} 
Nadaraya, E. A. (1964). On estimating regression. \textit{Theory of Probability and Its Applications}, \textbf{9}(1): 141--142. URL~\url{https://doi.org/10.1137/1109020}

\bibitem[Oppenheim, Willsky, and Nawab(1997)]{Oppenheim:1997}
Oppenheim, A.~V., Willsky, A.~S., and Nawab, S.~H.~(1997). \textit{Signals and Systems} (2nd ed.). Prentice Hall.

\bibitem[Parzen(1962)]{Parzen:1962} 
Parzen, E.~(1962). On estimation of a probability density function and mode. \textit{Annals of Mathematical Statistics}, \textbf{33}(3), 1065--1076.

\bibitem[R Core Team(2025)]{R:2025}
R Core Team (2025). \textit{R: A language and environment for
statistical computing}. R Foundation for Statistical Computing,
Vienna, Austria. URL~\url{https://www.R-project.org/}. 

\bibitem[Rosenblatt(1956)]{Rosenblatt:1956} 
Rosenblatt, M.~(1956). Remarks on some nonparametric estimates of a density function. \textit{Annals of Mathematical Statistics}, \textbf{27}(3), 832--837.

\bibitem[Rudemo(1982)]{Rudemo:1982} 
Rudemo, M.~(1982). Empirical choice of histograms and kernel density estimators. \textit{Scandinavian Journal of Statistics}, \textbf{9}(2), 65--78.

\bibitem[Sch\"{o}lkopf et al(1999)]{Scholkopf:1999} 
Sch\"{o}lkopf, B., Williamson, R.~C., Smola, A.~J., Shawe-Taylor, J., and Platt, J. C.~(1999). Support vector method for novelty detection. \textit{Advances in Neural Information Processing Systems}, \textbf{12}: 582-588.

\bibitem[Sheather and Jones(1991)]{Sheather:1991} 
Sheather, S. J., \& Jones, M. C. (1991). A reliable data-based bandwidth selection method for kernel density estimation. \textit{Journal of the Royal Statistical Society, Series B}, \textbf{53}(3): 683--690. 


\bibitem[Schuster(1985)]{Schuster:1985}
Schuster, E.~F.~(1985). Incorporating support constraints into nonparametric estimators of densities. \textit{Communications in Statistics—Theory and Methods}, \textbf{14}(5): 1123--1136.


\bibitem[Silverman(1982)]{Silverman:1982} 
 Silverman, B.~W.~(1982). Algorithm AS 176: Kernel density estimation using the fast Fourier transform. \emph{Journal of the Royal Statistical Society, Series C (Applied Statistics)}, \textbf{31}(1): 93--99.

\bibitem[Silverman(1986)]{Silverman:1986} 
Silverman, B.~W.~(1986). \textit{Density Estimation for Statistics and Data analysis}. Chapman and Hall.


\bibitem[Sturges(1926)]{Sturges:1926} 
Sturges, H.~A.~(1926). The choice of a class interval. \textit{Journal of the American Statistical Association}, \textbf{21}: 65--66.

\bibitem[Sutton and Barto(2018)]{Sutton:2018} 
Sutton, R. S., \& Barto, A. G. (2018). \textit{Reinforcement learning: An Introduction} (2nd ed.). MIT Press.

\bibitem[Terrell and Scott(1992)]{Terrell:1992} 
Terrell, G.~R.~and Scott, D.~W.~(1992). Variable kernel density estimation. \textit{Annals of Statistics}, \textbf{20}(3), 1236--1265.

\bibitem[Tsybakov(2009)]{Tsybakov:2009}
Tsybakov, A.~B.~(2009). \textit{Introduction to nonparametric estimation}. Springer.

\bibitem[Venables and Ripley(2002)]{Venables:2002}
Venables, W.~N.~and Ripley, B.~D.~(2002). \textit{Modern Applied Statistics with S}. New York: Springer.

\bibitem[Wand and Jones(1995)]{Wand:1995} 
Wand, M.~P.~and Jones, M.~C.~(1995). \textit{Kernel Smoothing}. \textit{CRC Press}.





\end{thebibliography}

\vspace{.3in}
\noindent
\begin{thebibliography}{9}
\bibitem[David and Nagaraja(2003)]{David:2003}
David, H.~A.~and Nagaraja, H.~N.~(2003). \textit{Order statistics}, 3rd ed.~Wiley.

\end{thebibliography}
\end{document}